\PassOptionsToPackage{dvipsnames}{xcolor}
\documentclass[letterpaper,twocolumn,10pt]{article}
\usepackage{usenix2019_v3}

\usepackage{tikz}

\usepackage[english]{babel}
\usepackage{blindtext}
\usepackage[vlined,linesnumbered,ruled]{algorithm2e}
\usepackage{bbding, enumitem}
\usepackage{makecell, booktabs, threeparttable, balance, multirow, multicol, booktabs, amsmath, amssymb, pifont, anyfontsize, fontawesome5}

\newcommand{\tabincell}[2]{\begin{tabular}{@{}#1@{}}#2\end{tabular}}
\newcommand{\ie}{\emph{\textsf{i.e.,}}\xspace}
\newcommand{\eg}{\emph{\textsf{e.g.,}}\xspace}
\newcommand{\first}{\textsf{(i)}\xspace}
\newcommand{\second}{\textsf{(ii)}\xspace}
\newcommand{\third}{\textsf{(iii)}\xspace}
\newcommand{\fourth}{\textsf{(iv)}\xspace}

\newcommand{\parab}[1]{\vspace{0.01in}\noindent{\bf #1} }

\newcommand{\g}[1]{\small\textcolor[rgb]{0, 0.5, 0}{#1}}

\newcommand{\idp}{\textsf{INDP}\xspace}
\newcommand{\sys}{\textsf{BoS}\xspace}
\newcommand{\imis}{\textsf{IMIS}\xspace}

\begin{document}
\pagenumbering{gobble}
\date{}

\title{Brain-on-Switch: Towards Advanced Intelligent Network Data Plane via \\ NN-Driven Traffic Analysis at Line-Speed}

\author{
{\rm Jinzhu Yan\textsuperscript{1,$*$}}
\quad
{\rm Haotian Xu\textsuperscript{1,$*$}}
\qquad
{\rm Zhuotao Liu\textsuperscript{1,2,~\faIcon{envelope}}} 
\qquad
{\rm Qi Li\textsuperscript{1,2}}
\qquad
{\rm Ke Xu\textsuperscript{1,2}} \\
{\rm Mingwei Xu\textsuperscript{1,2}}
\qquad
{\rm Jianping Wu\textsuperscript{1,2}}\\
{\rm \textsuperscript{1}}~Tsinghua University
\qquad
{\rm \textsuperscript{2}}~Zhongguancun Laboratory
}

\maketitle

\def\thefootnote{*}\footnotetext{Equal contribution. \quad \textsuperscript{\faIcon{envelope}} Corresponding author.}
\def\thefootnote{\arabic{footnote}}

\begin{abstract}

The emerging programmable networks sparked significant research on Intelligent Network Data Plane (\idp), which achieves learning-based traffic analysis at line-speed. 
Prior art in \idp focus on deploying tree/forest models on the data plane. 
We observe a fundamental limitation in tree-based \idp approaches: 
although it is possible to represent even larger tree/forest tables on the data plane, the flow features that are computable on the data plane are fundamentally limited by hardware constraints. 
In this paper, we present \sys to push the boundaries of \idp by enabling Neural Network (NN) driven traffic analysis at line-speed. Many types of NNs (such as Recurrent Neural Network (RNN), and transformers) that are designed to work with sequential data have advantages over tree-based models, because they can take raw network data as input without complex feature computations on the fly. 
However, the challenge is significant: the recurrent computation scheme used in RNN inference is fundamentally different from the match-action paradigm used on the network data plane. \sys addresses this challenge by \first designing a novel data plane friendly RNN architecture that can execute unlimited RNN time steps with limited data plane stages, effectively achieving line-speed RNN inference; and \second complementing the on-switch RNN model with an off-switch transformer-based traffic analysis module to further boost the overall performance. 
We implement a prototype of \sys using a P4 programmable switch as our data plane, and extensively evaluate it over multiple traffic analysis tasks. The results show that \sys outperforms state-of-the-art in both analysis accuracy and scalability. 

\end{abstract}

\section{Introduction}


 
The emerging programmable network hardware (\eg P4 switch~\cite{bosshart2014p4}, NetFPGA~\cite{NetFPGA} and SmartNIC~\cite{2018azure,Netronome,Mellanox}) sparked significant research on Intelligent Network Data Plane (\idp). 
Compared with other \emph{AI-assisted networking designs} which deploy learning models on either end-hosts (\eg congestion control \cite{abbasloo2020classic,yan2021acc}) or network control plane (including auxiliary servers) (\eg routing control \cite{liu2021drl,zhou2021primus}), 
\idp is \emph{forwarding-native} since it deploys learning models directly on network data plane. Thus, the key merit of \idp, as first summarized in~\cite{netbeacon}, is that it enables intelligent network traffic analysis at line-speed based on data-driven learning models rather than empirical rules/protocols. 

The initial exploration of \idp 
begins with extracting fine-grained flow information from the programmable data plane to support a variety of overarching applications, such as covert channel detection~\cite{xing2020netwarden}, RTT measurement~\cite{sengupta2022rtt}, traffic classification~\cite{barradas2021flowlens}, and DDoS mitigation~\cite{liu2021jaqen}. 
Yet, the subtle distinction between these early approaches and the native \idp paradigm is that they fail to directly deploy learning models on the data plane due to various hardware constraints. For example, the lack of support for floating-point arithmetic on P4 switches makes it significantly more difficult to execute model inference on the data plane than on general-purpose processors like CPUs and GPUs.

The community since then make substantial progress on realizing tree-based \idp~\cite{busse2019pforest, lee2020switchtree, zheng2021planter,xie2022mousika,netbeacon,zheng2022iisy,xavier2021programmable}, based on the insight that the decision making process in tree-models can be implemented using match-action tables on the programmable data plane. State-of-the-art (SOTA) in this regard is NetBeacon~\cite{netbeacon} which designs a novel ternary table encoding mechanism to efficiently deploy fairly large tree/forest models on programmable switches. 
Further, the recent art~\cite{N3IC, siracusano2018network, siracusano2020running, sanvito2018can} embrace neural networks by deploying binarized Multi-Layer Perceptron (MLP) on SmartNIC. Yet, the capacity of SmartNIC (\eg \mbox{$2\times40$ GbE} for Netronome Agilio CX~\cite{Netronome}), which co-locates with an end-host, is several orders of magnitude smaller than the in-network programmable switches (\eg \mbox{6.4 Tbps} for Barefoot \mbox{Tofino 1} switch). 

In this paper, we propose Brain-on-Switch (\sys) to advance state-of-the-art of \idp in two fundamental ways. 
First, \sys enables the use of Neural Network (NN) in \idp. NNs have several advantages over tree-based models for  traffic analysis. For instance, Recurrent Neural Network (RNN), a type of NN designed to work with sequential data, outperforms tree-models in both efficiency (\eg not requiring complicated feature computations on the fly, consuming fewer hardware resources to maintain per-flow state, etc.) and accuracy (especially when handling more complex tasks, such as \emph{multi-class} traffic classification). 
Second, \sys is architecturally complete in the sense that it can accommodate full-precision and advanced models in \idp. Hardware limitations (\eg lack of floating-point number support) on switches force model binarization~\cite{alizadeh2019empirical, xie2022mousika, N3IC}, which, unfortunately, reduces performance. 
Although prior art (\eg IIsy~\cite{zheng2022iisy}, \cite{silla2011survey}) mentioned the hybrid analysis concept of forwarding certain flows to large tree-based models deployed at the endpoint for reevaluation, they lack the fundamental design to precisely control the amount of such \emph{escalated} flows processed off-switch. In contrast, \sys proposes a novel approach to accommodating advanced off-switch models (\eg transformer-based models) into \idp to improve the overall analysis performance, while ensuring that the vast majority of traffic (\eg over 95\% of flows) is still analyzed at line-speed on the data plane. 

Concretely, \sys has the following innovative designs: 

\first A novel binary RNN architecture that retains full-precision model weights during on-switch inference (\ie only activation functions are binarized), realized by encoding the complex layer forward propagation functions as match-action tables. Compared to the fully-binarized MLP model~\cite{N3IC}, our binary RNN exhibits substantial performance advantages. 

\second A sliding-window based computation scheme to execute unlimited RNN time steps using limited forwarding stages on switches. \sys overcomes various switch hardware limitations in realizing the critical operations essential to this computation scheme, such as the read/write of a ring buffer like data structure, and an \textsf{argmax} like operation to make comprehensive inference decisions by aggregating the intermediate analysis results as a flow proceeds.  

\third An analysis-escalation module to accommodate full-precision transformer-based models in \sys. The key design is two-fold: accurately identifying the flows for which on-switch analysis confidence is insufficient, and designing an Integrated Model Inference System (\imis) to enable fast off-switch inference for escalated flows.


\parab{Contributions.} 
The major contribution of this paper is the design, implementation and evaluation of \sys, the first \idp design that enables NN-driven traffic analysis at line-speed. We implement a prototype of \sys and evaluate it extensively using several use cases. The experimental results show that  \sys outperforms SOTA in analysis accuracies by non-trivial margins, achieving up to ${\sim}19\%$ higher F1-scores than tree-based NetBeacon~\cite{netbeacon} and up to ${\sim}40\%$ higher than binary MLP based N3IC~\cite{N3IC}. 
We further perform thorough system-level evaluations, demonstrating that \sys is scalable to handle high network loads (flow concurrency), attributing to the co-design of the on-switch binary RNN and off-switch \imis. 
Finally, we evaluate hardware resource utilization by \sys. 


\section{Background and Motivation}
\label{sec:motivation}

\parab{Programmable Network Data Plane.}
The emerging of Protocol-Independent Switch Architecture (PISA) enables flexible data plane programmability, empowering fast innovations of networking designs.
In PISA, the switching pipeline can be programmed via P4~\cite{bosshart2014p4}, a domain-specific programming language. A PISA pipeline consists of a parser for header parsing, multiple match-action stages for header fields and metadata manipulating, and a deparser for header reassembling. In general, the  actual packet processing logic is implemented using these match-action stages. 
PISA also supports components for stateful storage, such as registers. 

Despite the programmability mentioned above, PISA has the following limitations. First, only simple operations like add, subtract, shift and bit-wise operations are supported, excluding floating numbers, multiplication, division and complex comparisons. Second, the resources (such as the number of stages, SRAM, TCAM) are limited. For instance, on Barefoot Tofino 1, each pipeline has 12 stages, \mbox{120 Mbit} SRAM and \mbox{6.2 Mbit} TCAM~\cite{netbeacon}. Finally, each register can only be accessed once through an atomic operation for each packet. 

\begin{figure*}[t]
    \centering
    \includegraphics[width=\textwidth]{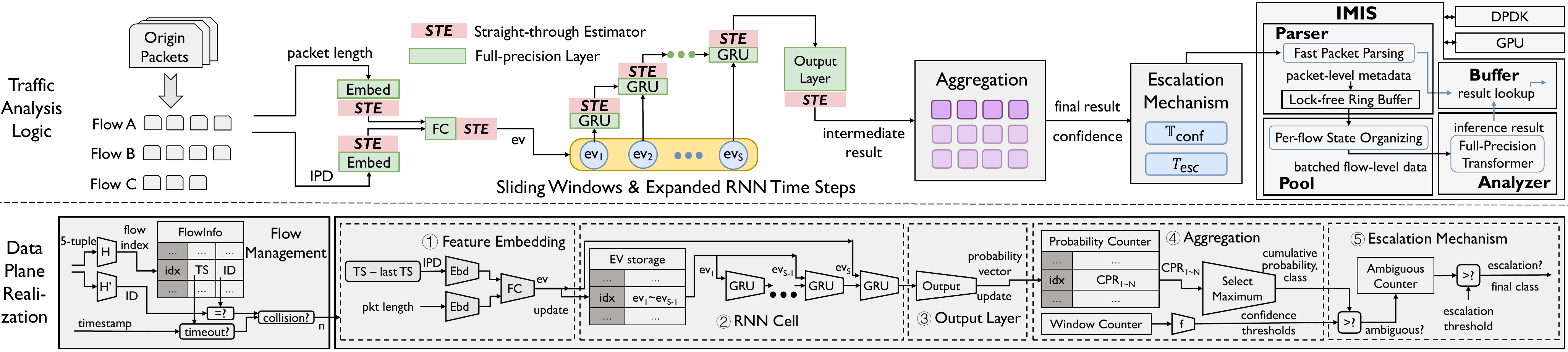}
    \caption{\label{fig:overview} The \sys architecture enables NN-driven traffic analysis in \idp.}
\end{figure*}

\parab{RNNs and Transformers.}
RNN~\cite{RNN} is designed to process sequence data of varying lengths by maintaining an internal state (\ie the hidden state). 
Specifically, given the input $x_t \in \mathbb{R}^m$ at time step $t$, the hidden state $h_t \in \mathbb{R}^n$ is calculated as $h_t = {\rm tanh}(W[x_t, h_{t-1}] + b)$, where $W$ and $b$ are trainable parameters. As $h_t$ encodes the current input $x_t$ and the historical information $h_{t-1}$ at the same time, RNN can capture the relationships between the data points in a sequence. 
The algorithm used to calculate the hidden states is called a recurrent unit, and the two most popular recurrent units are LSTM~\cite{LSTM} and Gated Recurrent Unit (GRU)~\cite{GRU}.


Transformers~\cite{vaswani2017attention} excel at modeling sequential data. 
Recently, several traffic analysis approaches~\cite{lin2022bert, yatc, lin2023novel, xie2023rosetta, deng2023robust} treat the bytes of packets as words or images, and introduce a variety of transformer-based models to achieve impressive traffic classification performance. In addition, the self-supervised pre-training paradigm used in transformer-based traffic analysis requires a small amount of labeled data.

\parab{Motivation.} 
Our community make substantial progress~\cite{busse2019pforest, lee2020switchtree, zheng2021planter,xie2022mousika,netbeacon, zheng2022iisy, xavier2021programmable} on realizing Intelligent Network Data Plane (\idp) by embedding decision tree models in the forwarding pipeline of programmable switches. Their key insight is that the decision making process in tree/forest models (\ie comparing a value to a threshold and then moving on to the next tree node until a leaf node is reached) is very similar to the match-action table paradigm used on the data plane. 
For instance, state-of-the-art NetBeacon~\cite{netbeacon} designs a novel coding algorithm to effectively represent multiple tree models using ternary matching tables. 

We forecast a performance ceiling in further innovating tree-based \idp designs. 
Specifically, tree models often rely on advanced feature engineering (\ie extracting various types of statistics/properties/attributes from raw data) to boost accuracy. However, the features that are computable on the data plane at line-speed are fundamentally limited due to hardware constraints. 
For instance, flow features such as the s.t.d., frequency, and percentile of packet lengths are critical to tree models~\cite{netbeacon,rahbarinia2013peerrush}. Yet, computing these features is either impossible or difficult, often requiring ad-hoc tricks to estimate these statistics. 
For instance, prior art~\cite{netbeacon} estimates s.t.d. of packet lengths upon receiving certain packets (\ie the $2^k$-th packet in each flow), indicating it can only execute inference at these locations. The limitation is obvious: an inference error obtained on the $2^k$-th packet cannot be corrected until the arrival of the $2^{k{+}1}$-th packet. 

\parab{Design Goals.}
Therefore, philosophically, it is worth asking: can we expand the boundaries of \idp to a new type of learning models that is not limited by the availability of flow features on the data plane. 
In this paper, we address this research question concretely by enabling NN-driven traffic analysis in \idp. Unlike tree/forest models, many types of NNs (such as RNNs and transformers) that are designed to process sequential data can directly take raw network traffic data as input, eliminating the requirements of computing complex features on the data plane on the fly. 
However, the incorporation of NNs into \idp presents significant challenges. For instance, the recurrent computation scheme in RNN is fundamentally different from the match-action paradigm on the data plane, making it more difficult to realize on-switch RNN inference. Additionally, existing transformer-based traffic analysis approaches~\cite{lin2022bert, yatc, lin2023novel, xie2023rosetta, deng2023robust} simply treat network traffic as another form of sequential data, without constructing appropriate systems to analyze the network flows online while they are traversing the data plane. 

To address these challenges, we architect \sys, the first NN-driven \idp system. 
A recent art N3IC~\cite{N3IC} explores to deploy binary MLP models on SmartNIC, which is more computationally flexible, yet with much lower throughput than programmable switches. 
We focus on programmable switch based \idp in this paper (although we also compare the traffic analysis accuracy of \sys with N3IC in our evaluations). 
Concurrent with \sys, Broadcom unveils the early-stage development of their novel NN inference switching chip~\cite{netgnt}, underlining the significance of \idp.

\section{\label{sec:overview} Design Overview}

We plot the architecture of \sys in Figure~\ref{fig:overview}. The overarching traffic analysis logic in \sys centers around \first a data plane friendly RNN inference architecture and \second a co-design with an Integrated Model Inference System (\imis) to accommodate full-precision transformer-based traffic analysis models. 
The key designs toward hardware friendliness are two-fold: \first realizing the online forward propagation of RNN layers via offline-trained input-output-mapping tables, and \second executing unlimited recurrence of RNN time steps via a sliding window mechanism that recurrently processes fixed-length packets segments.   
The key to co-design with \imis is accurately identifying the flows that do not receive sufficient confidence from the on-switch analysis, and to only escalate these flows to the off-switch \imis, so that \sys still processes the vast majority of traffic on-switch (\eg over 95\% flows). 
Nevertheless, 
we optimize the system design of \imis so that a single instance of \imis can process ten million packets per second while maintaining low inference latencies (see \S~\ref{subsec:deepdive}). 
This ensures the off-switch \imis is unlikely to be the bottleneck of \sys.

\section{Data Plane Friendly RNN Architecture}


\vspace{-0.2cm}
\subsection{Raw Packet Sequences as Input Features}

The on-switch RNN uses raw flow sequences as the input features, \ie the packet length sequence and the inter-packet delay (IPD) sequence. When a packet in the flow arrives at the switch, we extract its length and get IPD based on the subtraction of timestamps. Through feature embedding, these metadata are mapped into an embedding vector, which is stored in a sequence for subsequent model inference. 

Using raw sequences as input features has several key advantages over using statistical flow features (such as the mean and s.t.d. of packet lengths). First, the availability of critical flow features is greatly limited on switch (explained in  \S~\ref{sec:motivation}). 
Second, storing per-flow statistical features on switch is expensive: NetBeacon~\cite{netbeacon} consumes 2.5x storage to store features as evaluated in \S~\ref{subsec:end-to-end}. 
Finally, feature engineering, without careful designs, could result in overfitting problem. For instance, we notice that some features (\eg the number of packets with packet size in $[48,64)$) are heavily task-specific~\cite{netbeacon, barradas2018effective}, and features like port number can lead to overfitting on host configurations. 

\vspace{-0.3cm}
\subsection{Binary RNN Architecture} 

\begin{figure}[t]
    \centering
    \includegraphics[width=0.95\linewidth]{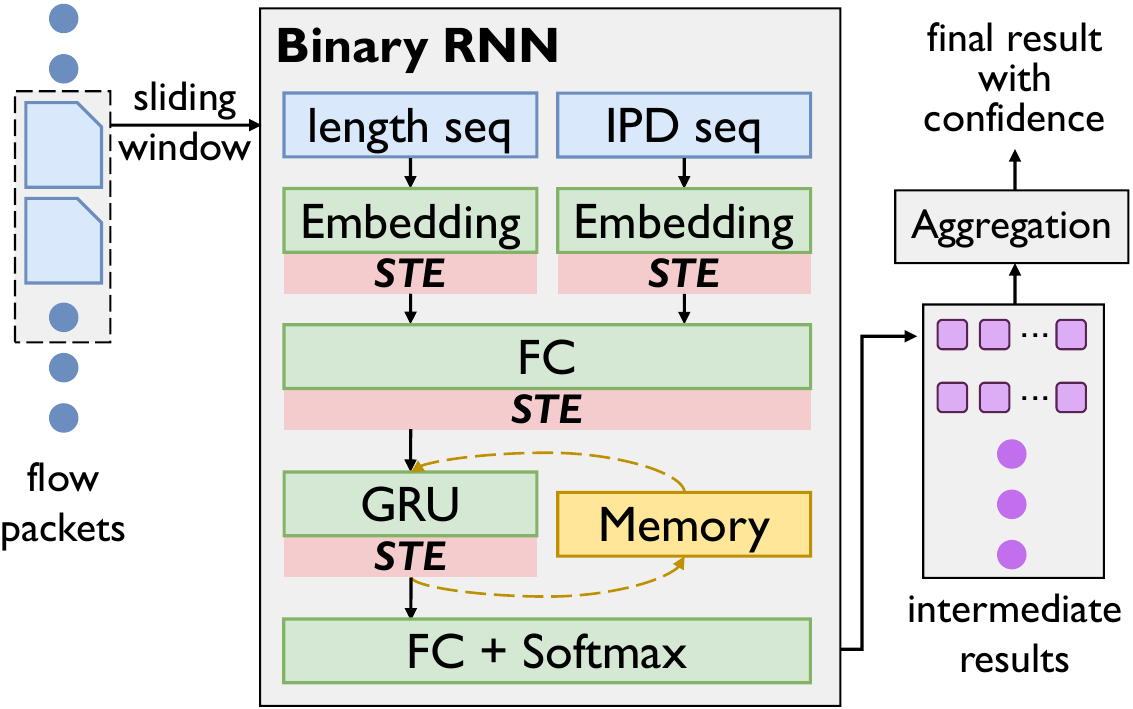}
    \caption{\label{fig:model} Data-plane-friendly binary RNN architecture.}
\end{figure}

As shown in Figure~\ref{fig:model}, architecturally, the binary RNN model in \sys consists of three building blocks: feature embedding, RNN cell and output layer. In the feature embedding, the length and IPD of each packet are passed through two different embedding layers, respectively, and then fed into a fully-connected layer to obtain an embedding vector. Taking the embedding vector sequence of per-flow packets as input, the RNN cell performs sequence analysis based on GRU~\cite{GRU}. In each time step of GRU calculation, the current embedding vector and the previous hidden state (\ie output activations of GRU) are used as input, and the output is used to update the hidden state. Finally, the hidden state of the last step is passed through the output layer (\ie a fully-connected layer with softmax) to obtain the analysis result. 

Each GRU calculation contains 3 Hadamard product operations and 3 multiplications with nonlinear functions, 
which cannot be natively implemented on programmable switches due to hardware constraints. To cope with this challenge, we perform binarization on neural network activation functions to enable  hardware-friendly model deployment. Specifically, we set all activation functions in the feature embedding and the RNN cell to Straight-Through Estimator (STE)~\cite{yin2019understanding}. STE performs a sign function in forward propagation, which makes all neural network activations +1 or -1. And in backward propagation, STE estimates the incoming gradient to be equal to the clipped outgoing gradient.

Prior art N3IC~\cite{N3IC} performs binarization on both weights and activations of an MLP model, and then implements fully-connected layer forward propagation on the SmartNIC using XOR and customized population count (\textsf{popcnt}) operations. The \textsf{popcnt} operation, unfortunately, is not friendly to the switch pipeline: 
realizing a single \textsf{popcnt} operation for a 128-bit string takes 14 switch stages. Yet, one 128bit-to-64bit fully-connected layer in an MLP model requires 64 \textsf{popcnt} operations. 
More crucially, as evaluated in \S~\ref{subsec:end-to-end}, full model binarization results in significant performance degradation. In Table~\ref{table:BRNNvsBMLP}, we summarize the key differences between our binary RNN and the binary MLP in N3IC~\cite{N3IC}. 





\begin{table}[t]
\centering
\caption{Binary RNN v.s. Binary MLP}
\label{table:BRNNvsBMLP}
\resizebox{\linewidth}{!}{
\begin{threeparttable}
\begin{tabular}{ccccc}
    \toprule
    
    Prior Work & \tabincell{c}{Binary\\Activations}  & \tabincell{c}{Full Precision\\Weights} & \tabincell{c}{Stage\\Consumption$\star$} & \tabincell{c}{Model\\Accuracy$\dag$} \\
    
    \midrule

    Binary MLP (N3IC\cite{N3IC}) & \CheckmarkBold & \XSolidBrush & High & Low \\
    
    Binary RNN & \CheckmarkBold & \CheckmarkBold & Low & High \\
    \bottomrule
\end{tabular}

\begin{tablenotes}
    \item[$\star$] Estimated if we were to implement the binary MLP on a programmable switch. 
    \item[$\dag$] See \S~\ref{subsec:end-to-end} for quantitative results. 
\end{tablenotes}

\end{threeparttable}
}

\end{table}

\subsection{Data Plane Native Model Inference} \label{subsec:online-inference}

We now present the data plane native RNN inference. 


\parab{Forward Propagation.} The key to retain full precision model weights in our RNN models is to avoid direct computations of the layer forward propagation on the data plane. To this end, \sys realizes forward propagation based on match-action table lookup. Specifically, since all activations are binarized to +1 or -1, the input and output vectors of any neural network layer (\eg the embedding layer, FC layer and GRU layer in Figure~\ref{fig:model}) are essentially bit strings. 
Therefore, regardless of what computations are executed in a neural network layer, we can realize \emph{equivalent input-output-relationship} by recording an enumerative mapping from input bit strings to output bit strings as a match-action table.  
Thus, in the online forward propagation through any layer, \sys uses input bit string as the key to match the output bit string stored in the corresponding table on switch. 
The caveat of this design is that the number of required entries $N$ in each table is determined by the number of input bits, \ie $N = 2^{\textsf{input bit-length}}$. We recognize this constraint and demonstrate, via experiments (\S~\ref{subsec:end-to-end}), that \sys can deploy efficient RNN models under this constraint. 

\begin{figure}[t]
    \centering
    \includegraphics[width=\linewidth]{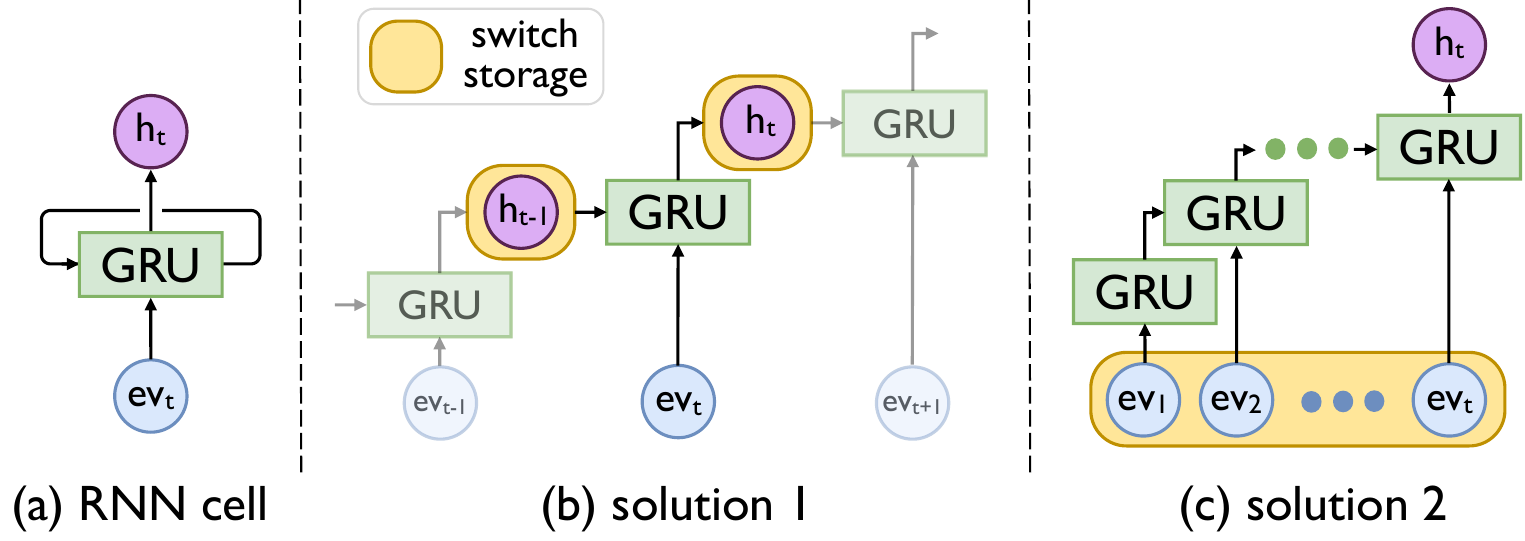}
    \caption{\label{fig:RNNtimesteps} The design choices for RNN time steps.}
\end{figure}

\parab{RNN Time Steps.} As shown in Figure~\ref{fig:RNNtimesteps}(a), a straightforward way to implement RNN time steps is to store the RNN hidden state for each flow sequence. Upon packet arrival, we read the previous RNN hidden state, perform layer forward propagation, and then update the hidden state. Unfortunately, due to hardware constraint, each register can only be accessed once when a packet traverses the switching pipeline. Therefore, we need to expand RNN time steps in serial stages, as shown in Figure~\ref{fig:RNNtimesteps}(b), where the read/write of hidden states spread across multiple stages. 
Alternatively, we can store a sequence of embedding vectors corresponding to packet sequence of a flow, as shown in Figure~\ref{fig:RNNtimesteps}(c). In online forward propagation, \sys calculates the embedding vector for each packet, updates the sequence in storage and executes all RNN time steps for this flow sequence in serial stages. We adopt this solution in the final prototype as it consumes fewer hardware resources. 

\parab{Sliding Window Mechanism.} 
Since the number of switch stages is limited, expanding the RNN time steps into serial stages would also limit the total number of time steps executable in our model. 
To address this problem, \sys designs a novel sliding window mechanism that can recurrently apply RNN inference on fixed-length flow segments. Therefore, the number of RNN time steps executable on a flow is no longer limited by switch stages. 
Specifically, in online traffic analysis, \sys uses a window with fixed-size $S$ to extract a segment of $S$ packets from the flow, executes $S$ RNN time steps on this segment to obtain an intermediate inference result $r_i$, and swifts the window by one packet to obtain a new segment, and repeats the process. 
Therefore, \sys can continuously execute RNN time steps as the flow proceeds. 


The key to fulfill the sliding window design is to properly aggregate these intermediate results. 
Specifically, upon receiving the $j^\text{th}$ packet, suppose that the binary RNN has processed $g$ full segments. Then the inference result for the $j^\text{th}$ packet shall consider all the $g$ intermediate inference results. 
In the case of multi-class traffic classification, one simple strategy is to select the majority class from these intermediate results. 
More crucially, we can co-design the aggregation algorithm with an off-switch module to improve the overall traffic analysis accuracy, as described below.



\vspace{-0.2cm}
\subsection{Analysis Escalation}
\label{subsec:ana_escalation}

Although \sys primarily relies on binary RNN to ensure line-speed traffic analysis, we still want to embrace full-precision and more advanced models (\eg transformers) to handle corner cases.  
For instance, in multi-class traffic classification, it is possible that none of the classes dominates (\eg the numbers of flow segments for different classes are close to each other). In this case, adopting the majority voting policy may reduce classification confidence. 

To this end, \sys adopts an off-switch traffic analysis module co-located with the programmable data plane. We recognize two challenges in accommodating this analysis module. First, the aggregation algorithm must be carefully designed to ensure that it can accurately capture ambiguity, while avoiding consistently escalating flows (\ie only a small fraction of flows should be escalated to the co-located analysis module in order to preserve line-speed analysis for the vast majority of traffic). Second, the analysis module adopts a transformer-based model to improve accuracy. However, due to the complex computations required for inference, it is non-trivial to scale the online analysis throughput to maintain the high speed of network forwarding. 

\parab{The Escalation Mechanism.} To address the first challenge, we measure the \emph{classification confidence} in the aggregation algorithm. Specifically, for each extracted packet segment in a flow, the binary RNN predicts an intermediate inference result, which is a vector of probabilities, one for each class. Suppose that upon receiving the $j^\text{th}$ packet of the flow, the binary RNN has processed $g$ packet segments for the flow (\ie the arrival of the $j^\text{th}$ packet will form the $\{g{+}1\}^\text{th}$ segment). Then, our algorithm aggregates all $g+1$ intermediate inference results by accumulating the per-class prediction probabilities. The class with the largest cumulative probability is selected as the inference result for the $j^\text{th}$ packet. 

Whether a flow should be escalated is determined by the number of \emph{ambiguous packets} in the flow. Upon receiving the $j^\text{th}$ packet, suppose the largest cumulative probability among all classes is $\textsf{CPR}_m$ and the total number of intermediate results is $\textsf{wincnt}$, then the classification confidence for the $j^\text{th}$ packet is quantified as $\textsf{CPR}_m / \textsf{wincnt}$. The packet is considered ambiguous if its confidence is below a predefined \emph{confidence threshold}. We use $\mathbb{T}_\textsf{conf}$ to represent the vector of confidence thresholds, one for each class. The flow is escalated when the number of ambiguous packets in the flow exceeds a predefined \emph{escalation threshold} $T_\textsf{esc}$.

\begin{figure}[t]
    \centering
    \includegraphics[width=\linewidth]{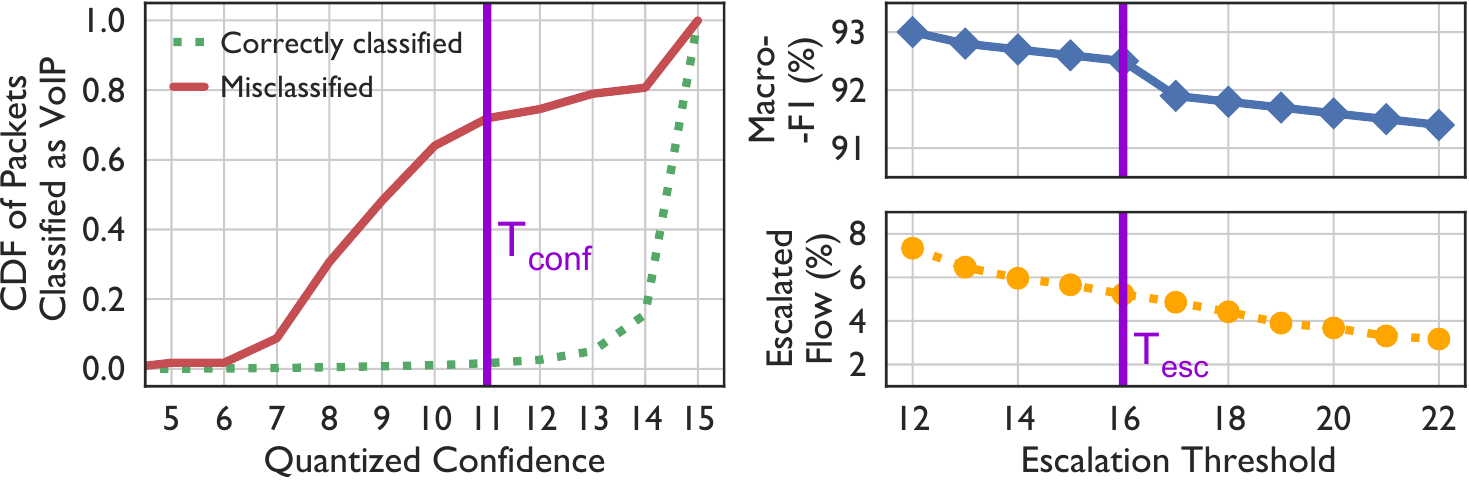}
    \caption{\label{fig:thresholds} The selection of $T_\textsf{conf}$ and $T_\textsf{esc}$.}
\end{figure}

$\mathbb{T}_\textsf{conf}$ and $T_\textsf{esc}$ are learned based on the distributions of the classification confidences of the training samples. Consider the example in Figure~\ref{fig:thresholds}. For the VoIP class in the ISCXVPN2016 dataset (see detailed descriptions in \S~\ref{subsec:experiment_setup}), we plot the CDF of the confidence scores for both correctly classified packets and misclassified packets. The confidence scores are quantized because they are eventually computed on the data plane (see \S~\ref{subsubsec:decision_making}). An appropriate $\mathbb{T}_\textsf{conf}$ should escalate as many misclassified packets as possible without affecting correctly classified packets. To this end, we design the following loss function to train our binary RNN. 

Suppose $p_i$ is the RNN's prediction probability for class $i$ and $y$ is the ground-truth class. The classic cross entropy (CE) loss is ${\rm CE} = -\log(p_y)$. The CE loss solely considers to improve the model's ability to predict the correct class. Our loss is defined as $\mathcal{L}_{1} = -(1-p_y)^\gamma\log(p_y) - \lambda \sum_{i\neq y} p_{i}^\gamma \log (1-p_i)$, 
which adds another item to explicitly negate the model's prediction on the non-ground-truth classes. The factor $\lambda$ balances the two items, while the modulating factors $(1 - p_y)^\gamma$ and $p_i^\gamma$ down-weight easy samples and focus on hard samples, as proposed in the Focal Loss~\cite{focalloss}. Intuitively, this loss enhances the confidence differences between misclassified and correctly classified packets by reducing $p_i(i\neq y)$ while retaining high $p_y$. 
Since our aggregation algorithm chooses the class with the largest accumulative probability, a simplified version of the above loss function is to only reduce the maximum prediction probability among all the non-ground-truth classes, \ie $\mathcal{L}_{2} = -(1-p_y)^\gamma\log(p_y) - \lambda p_\textsf{false}^\gamma \log (1-p_\textsf{false})$,
where $p_\textsf{false}$ is the largest $p_i$ among all the non-ground-truth classes. We thoroughly evaluate these loss functions in \S~\ref{subsec:deepdive}.

Once $\mathbb{T}_\textsf{conf}$ is determined, we set $T_\textsf{esc}$ to control the amount of escalated flows. As shown in Figure~\ref{fig:thresholds}, we select a $T_\textsf{esc}$ to ensure that no more than 5\% flows are escalated to the co-located analysis module for further analysis.

\parab{Integrated Model Inference System.} To address the second challenge, we design an \imis that enables fast online inference for a full-precision transformer-based model. As illustrated by Figure~\ref{fig:overview}, \imis orchestrates four types of stateful and single-threaded tasks (called \emph{engines}) to realize a \emph{non-blocking} traffic processing pipeline. We describe the transformer model training and the architecture of \imis in \S~\ref{sec:implementation}. 




\SetKwComment{Comment}{$\triangleright$~}{}
\SetKwInOut{KwIn}{Define}
\begin{algorithm}[t]
    \caption{Integrated Traffic Analysis Logic}
    \label{alg:workflow}
    \small
    \DontPrintSemicolon
    \SetAlFnt{\small}

    \KwIn {$\textsf{WIN}[0 \dots S{-}1]$ sliding window; $N$ No. of classes; $\textsf{CPR}[0 \dots N{-}1]$ per-class results; $\mathbb{T}_\textsf{conf}[0 \dots N{-}1]$ the per-class confidence threshold; $T_\textsf{esc}$ the escalation threshold}


        \If {{\rm \textsf{{FlowManager}}}({\rm \textsf{packet}} $\mathcal{P}$) fails} {
            Fall back to use the per-packet model, and exit 
        }
        Retrieve the flow state for $\mathcal{P}$  \;
        
        \If {$\mathcal{P}$ is matched by the $\textup{\textsf{EscTable}}$} {
            Forward $\mathcal{P}$ to \imis, and exit \Comment*{\g{Escalated flows}}
        }

        $\textsf{pktcnt} \gets \textsf{pktcnt} + 1$ \Comment*{\g{Count packets}}

        $ev \gets \textsf{\textbf{FeatureEmbedding}} (\mathcal{P}.\textsf{length}, \mathcal{P}.\textsf{IPD})$ \;

        $\textsf{WIN}[\textsf{pktcnt} ~\%~ S] \gets ev$ \Comment*{\g{Slide the window}}
        
        \eIf (\Comment*[f]{\g{The first $S-1$ packets}}) {{\rm \textsf{pktcnt}} $< S$} {
            Pre-analysis packet handling \Comment*{\g{See \S~\ref{app:preanalysis}}} 
        } { 
            $h \gets \vec{0}$ \;
            
            \For (\Comment*[f]{\g{RNN time steps}}) {$i \gets 1 $ {\rm \textbf{to}} $S$} {
                $ev_i \gets \textsf{WIN}[(\textsf{pktcnt} + i) ~\%~ S]$ \;
                $h \gets \textsf{\textbf{RNNCell}}(ev_i, h)$
            }

            $PR \gets \textsf{\textbf{OutputLayer}}(h)$
            \Comment*{\g{Intermediate result: a probability vector}}
            
            $\textsf{CPR} \gets \textsf{CPR} + PR$ \;
            
            $\textsf{Class} \gets \textsf{argmax}(\textsf{CPR})$ \Comment*{\g{Measure confidence}} 

            $\textsf{wincnt} \gets \textsf{wincnt} + 1$ \Comment*{\g{No. of windows}}

            \If {$\textup{\textsf{CPR}}[\textup{\textsf{Class}}] < \mathbb{T}_\textsf{conf}[\textup{\textsf{Class}}] * \textup{\textsf{wincnt}}$} {
                $\textsf{esccnt} \gets \textsf{esccnt} + 1$ \Comment*{\g{Ambiguous packets}}
            }

            \If {{\rm \textsf{esccnt}}$~\geq~ T_\textsf{esc}$} {
                Initiate escalated analysis for subsequent packets
            }\label{line:escalation}
            
            \lIf{{\rm \textsf{pktcnt}}$~\%~ \mathcal{K} = 0$}{$\textsf{Reset}(\textsf{wincnt}, \textsf{CPR})$} \label{line:reset}
        }

\end{algorithm}

\subsection{Integrated Analysis Logic}
Algorithm~\ref{alg:workflow} summarizes the complete logic of our online traffic analysis in \sys. 
Because our binary RNN model leverages flow-level data for inference, \sys designs a dedicated flow manager to store per-flow state. When a packet $\mathcal{P}$ is received, the flow manager first checks if per-flow state for $\mathcal{P}$ has already been allocated. If not, the flow manager allocates new per-flow storage for the packet. The packet then enters the normal flow-aware inference pipeline based on the retrieved flow state. 
Due to the limited capacity of the switch, when the flow manager cannot allocate storage for a new flow, \sys falls back to analyzing the packets of that flow using a tree model trained only using per-packet features (\eg packet length, TTL, Type of Service, TCP offset). This tree model is deployed on the data plane alongside our binary RNN model. The detailed design of the flow manager is deferred to \S~\ref{app:flow_management}. 

We elaborate on one key design that has not been thoroughly discussed yet. In  line~\ref{line:reset}, we periodically reset the window counter and per-class results every $\mathcal{K}$ packets. 
This effectively clears the contributions of very ancient flow segments (\ie more than $\mathcal{K}+1$ packets apart) when aggregating the intermediate inference results. 
This design rationale is that if we obtain a sub-flow $f_\textsf{sub}$ by extracting a continuous and sufficiently long sequences of packets (starting from any position) from a flow $f$, it is very likely that $f_\textsf{sub}$ and $f$ are classified as the same class. Thus, clearing the results of very remote segments will not affect  traffic analysis results.
On the contrary, without periodical reset, the per-class results $\textsf{CPR}$ would be consistently accumulated. To prevent buffer overflow, we need to allocate more bits to store $\textsf{CPR}$, which, unfortunately, results in significant hardware resource consumption (see \S~\ref{subsubsec:decision_making}). 
Note that the periodical reset does not clear the embedding vectors for the previous $S{-}1$ packets. 



\vspace{-0.2cm}
\section{Model Realization on the Data Plane}



\subsection{Embedding Vector Storage and Retrieval}
\label{subsubsec:embedding_vector_storage} 

As shown in Figure~\ref{fig:RNNtimesteps}(c), the RNN cell takes the embedding vectors of the packets in each sliding window (with length $S$) as input. 
Consider the flow segment starting with the $k^\text{th}$ packet and ending with the $\{k{+}S{-}1\}^\text{th}$ packet. Upon arrival of the $\{k{+}S{-}1\}^\text{th}$ packet, the current flow segment is full and is ready to execute all $S$ RNN steps simultaneously, \ie using the embedding vector of each packet as the key to obtain the matched output in each GRU table ($S$ tables in total). 
Thus, before a segment is full, we need to temporally hold the embedding vectors for prior $S{-}1$ packets. 
We design a ring buffer with $S{-}1$ bins (registers) to store embedding vectors. In particular, the $k^\text{th}$ packet in the segment is stored in the $\{k ~\%~ (S{-}1){+}1\}^\text{th}$ bin (indexed from $1$) of the ring. The embedding vector of the $\{k{+}S{-}1\}^\text{th}$ packet eventually takes the bin occupied by the $k^\text{th}$ packet, which will be out-of-scope upon the arrival of the $\{k{+}S\}^\text{th}$ packet.  
The second benefit of the ring buffer design is that all bins are mutually independent, so that they can be accessed in parallel. 

\begin{figure}[t]
    \centering
    \includegraphics[width=\linewidth]{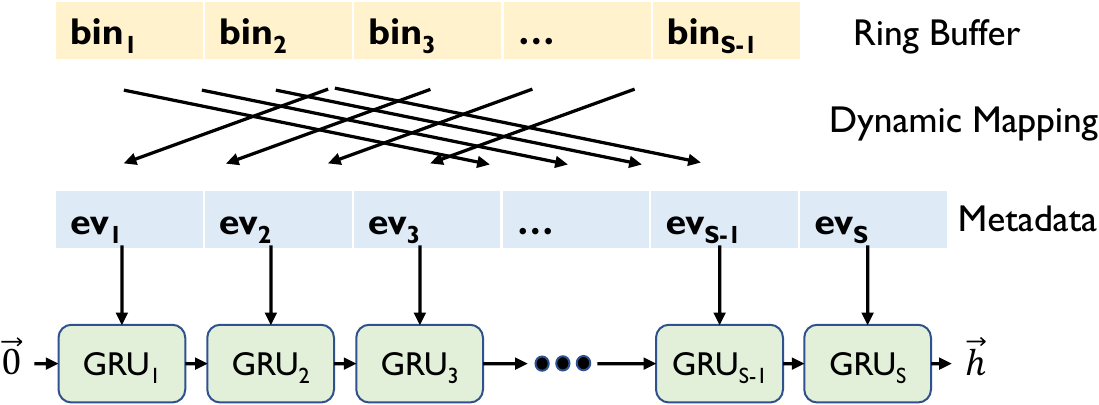}
    \caption{\label{fig:ev} Storage and  retrieval of embedding vectors.}
\end{figure}

The efficient storage of embedding vectors leads to a challenge in reading these vectors. 
Specifically, given a flow segment, the embedding vector of the first packet in this segment is not always stored in the first bin of the ring buffer. 
Therefore, it is incorrect to statically align the ring buffer with GRU tables, \ie using the value stored in the $k^\text{th}$ bin as the input key of the $k^\text{th}$ GRU table. 
Instead, the input of the $i^\text{th}$ GRU (for $i \in [1,S-1]$) should be read from the $\{(k-S+i)\%(S-1)+1\}^\text{th}$ bin. 
However, when declaring a lookup table for GRU on switch, the storage locations of its keys must be static and predetermined. Thus, to realize the above dynamic mapping, we need to first read values from the ring buffer to several intermediate variables (called metadata), and then dispatch the metadata to the proper GRU tables, as shown in Figure~\ref{fig:ev}. 

\subsection{Intermediate Results Aggregation }\label{subsubsec:decision_making} 

As described in \S~\ref{subsec:ana_escalation}, the key operation in our RNN inference is to select the largest cumulative probability from all intermediate inference results, \ie executing an \textsf{argmax} operation. 

\parab{Ternary-Matching Based Design.} 
\textsf{Argmax} is not a primitive available on the switch. We realize \textsf{argmax} based on an efficient data plane design of number comparison. Intuitively, number comparison can be accomplished by either conditional statements or exact-matching table matching. Neither of them, however, is scalable (see \S~\ref{subsec:number_comparison}). 

In \sys, we propose a scalable ternary-matching based design. 
Suppose \textsf{argmax} compares $n$ numbers each with $m$ bits. The key of a table entry consists $n$ segments, each with $m$ ternary bits (\ie 0, 1, or $\ast$). The value represents the winner (\ie the largest number).  
Starting from the most significant bit (MSB), to generate the $l^\text{th}$ bit for each key segment, there are $2^n$ possible cases. Consider a case $\mathcal{C}_{(l,k)}$ where the $l^\text{th}$ bits of the first $k$ ($k {\in} [1,n$-$1]$) segments are 1 and the $l^\text{th}$ bits of the remaining segments are 0. 
Clearly, the segments whose $l^\text{th}$ bits are 0 will not be the winner, so that we can stop further enumerating the lower bits (\ie the $\{l+1\}^\text{th}, \{l+2\}^\text{th}, ..., m^\text{th}$ bits) for these segments. Thus, among all $2^{(m-l) \cdot n}$ sub-cases of the case $\mathcal{C}_{(l,k)}$, we do not need further enumerations for $2^{(m-l)\cdot k}$ of them, achieving a $2^{(m-l)\cdot (n-k)}$ reduction ratio. 
Take $\mathcal{C}_{(1,1)}$ as an example: it represents the case where the first (\ie the most significant bit, MSB) of the first segment is 1 and the MSBs for other remaining segments are 0. Thus, all $2^{(m-1)n}$ sub-cases for $\mathcal{C}_{(1,1)}$ have clear winners and are collapsed into one key (\eg 1$\ast\ast\ast$,0$\ast\ast\ast$,0$\ast\ast\ast$ if $n{=}3,m{=}4$). 

Based on the above protocol, we derive the number of required table entries $F(n,m)$ as (see details in \S~\ref{appendix:argmax}). 
\begin{equation}
\begin{aligned}
F(n,m) &= 2*F(n,m-1) \\ 
&+ \sum_{i=1}^{n-1}\binom{n}{i}F(i,m-1)  \quad \textsf{for} \quad n, m\ge 2 \\
F(n,1) &= 2^n ~\textsf{for}~ n \ge 2; \quad F(1,m) = 1 ~\textsf{for}~ m \ge 1. 
\end{aligned}
\end{equation}

\begin{figure}[t]
    \centering
    \includegraphics[width=0.95\linewidth]{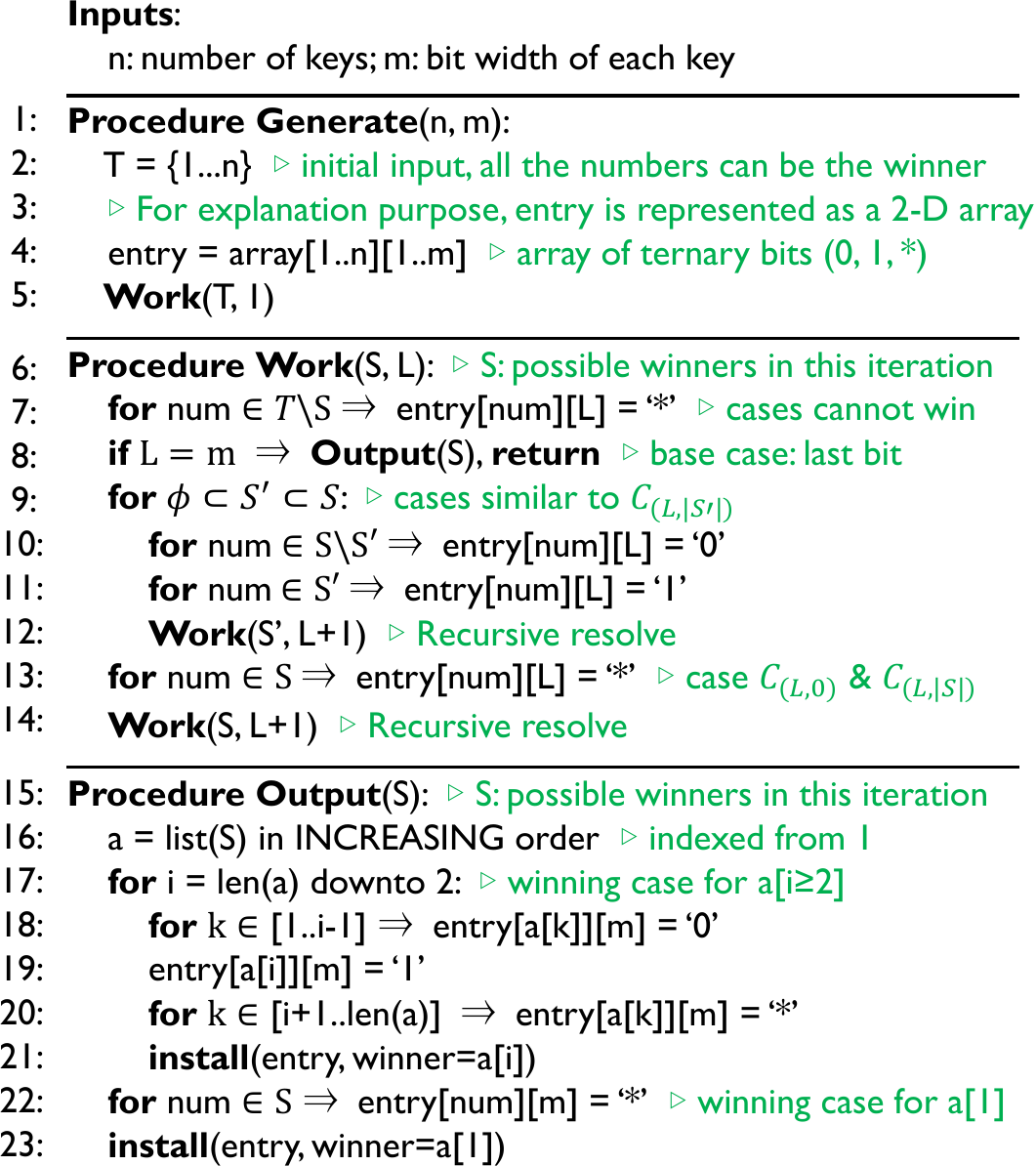}
    \caption{The procedure to generate a ternary-matching table to realize \textsf{argmax} on the data plane.
    }
    \label{fig:argmax}
\end{figure}

\parab{Further Optimizations.} 
We make two subsequent optimizations to further reduce $F(n,m)$. First, the two special cases $\mathcal{C}_{(l,n)}$ (\ie the $l^\text{th}$ bits in all $n$ segments are 1) and $\mathcal{C}_{(l,0)}$ (\ie the $l^\text{th}$ bits in all $n$ segments are 0) can be further merged. Specifically, for all $2^{m-l}$ sub-cases of $\mathcal{C}_{(l,n)}$, their  winners remain the same if we modify the $l^\text{th}$ bits of all n segments to 0; and similarly for $2^{m-l}$ sub-cases of $\mathcal{C}_{(l,0)}$, their winners remain the same if we modify the $l^\text{th}$ bits of all the n segments to 1. Thus, $\mathcal{C}_{(l,0)}$ and $\mathcal{C}_{(l,n)}$ can be merged by modifying the $l^\text{th}$ bit as an wildcard asterisk in each segment. 
We handle this merged case lastly  
in the current enumeration of the $l^\text{th}$ bit (see lines 13 and 14 in Figure \ref{fig:argmax}), so that these wildcard asterisks will not interfere with previous cases with higher priority (see lines 9 to 12 in Figure \ref{fig:argmax}).  
With this optimization, $F(n,m)$ is reduced as 
$F(n,m)=F(n,m-1)+\sum_{i=1}^{n-1}\binom{n}{i}F(i,m-1)$.

\begin{figure}[t]
    \centering
    \includegraphics[width=0.95\linewidth]{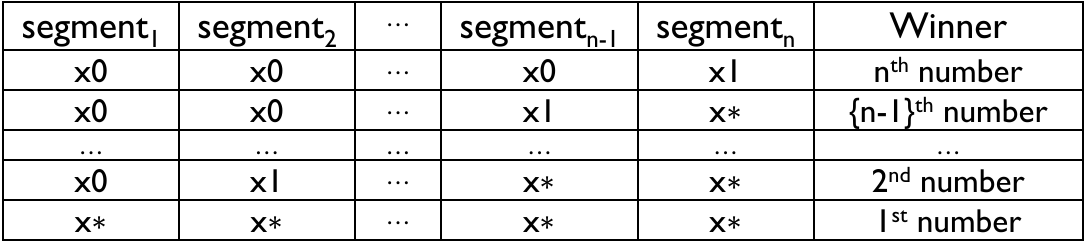}
    \caption{The reverse encoding for $F(n,1)$.}
    \label{fig:argmax_base_case}
\end{figure}

The second optimization is reducing the base case $F(n,1)$. By reversely encoding the one-bit number comparison (see Figure~\ref{fig:argmax_base_case}), $F(n,1)$ is reduced to $n$ from $2^n$. Combining both optimizations, we obtain $F(n,m) = nm^{n-1}$. 

\section{\label{sec:implementation} Implementation}

\begin{figure*}[t]
    \centering
    \includegraphics[width=\textwidth]{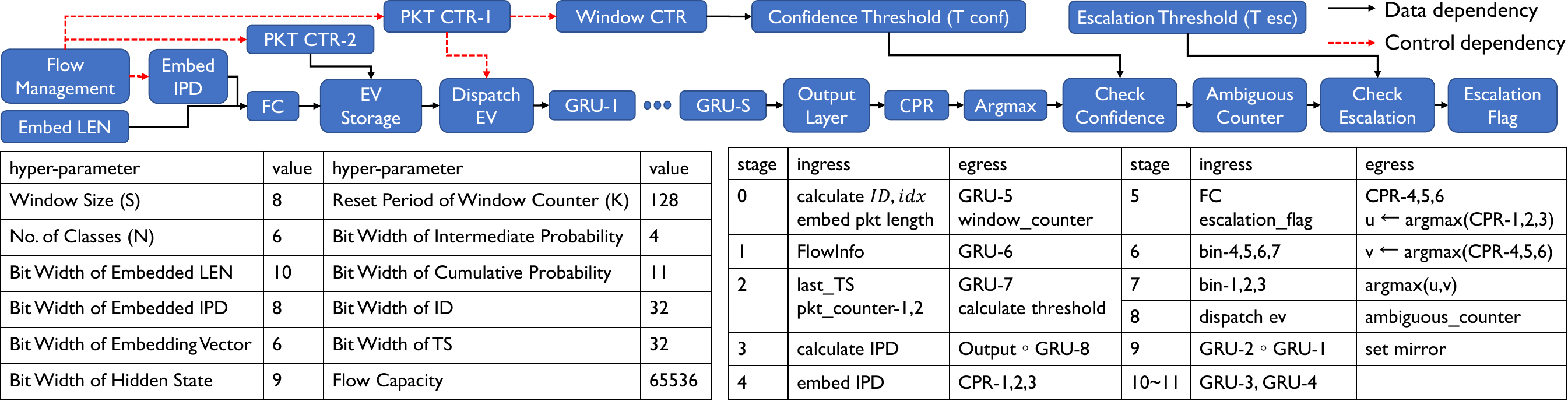}
    \caption{\label{fig:prototype} The breakdown of our on-switch RNN implementation on a Tofino programmable switch.}
\end{figure*}

Our \sys prototype\footnote{Available at https://github.com/InspiringGroup-Lab/Brain-on-Switch} includes: 1500 lines of Python code for model training, 1900 lines of P4 code for on-switch RNN, and 3300 lines of C++ code for \imis. To evaluate the prototype, an additional 1600 lines of code are developed. 


\parab{Model Training.} We train a binary RNN to analyze flow segments extracted by the sliding window. Given the window size $S$ and a  flow sample $(\mathcal{P}_1, \mathcal{P}_2, \dots)$ in the training dataset, we slice this flow into all possible packets segments (\eg consecutive $S$ packets like $(\mathcal{P}_1, \dots, \mathcal{P}_S)$ and $(\mathcal{P}_2, \dots, \mathcal{P}_{S+1})$) where the label of each segment is the flow label. 
For each segment, we use its packet length sequence and IPD sequence as inputs, and train the binary RNN to predict its label correctly. Recall that our binary RNN outputs vector of probabilities, one for each class. The training process is to maximize the prediction probability on the ground-truth class. 



We use YaTC~\cite{yatc}, a recent masked autoencoder (MAE)~\cite{he2022masked} based traffic transformer with multi-level flow representation, in \imis to analyze escalated flows. YaTC only uses the first 5 packets of a flow for analysis. For each packet, it extracts the first 80 header bytes and 240 payload bytes as inputs. We first determine the two thresholds in \S~\ref{subsec:ana_escalation} to collect the escalated flows in the training set, and then fine-tune the pre-trained YaTC model~\cite{yatc}  to obtain our final model.




\parab{On-Switch RNN Implementation.} We implement a prototype on our \mbox{Tofino 1} programmable switch. The top part of Figure~\ref{fig:prototype} shows the workflow of all the components in our prototype. The left-bottom table in Figure~\ref{fig:prototype} lists the hyperparameters of our prototype, and the right-bottom table lists the detailed per-stage arrangement of our components. Due to space constraints, we defer the detailed description of our prototype to \S~\ref{app:subsubsec:brnn}.
Although the hardware resources on the \mbox{Tofino 1} are very limited (\eg only 12 stages), we manage to implement a prototype that supports all four traffic analysis tasks evaluated in \S~\ref{subsec:experiment_setup}.
The on-switch RNN is programmable in runtime via the control plane (see \S~\ref{appendix:testbed}). 

\parab{\imis Implementation.} 
The core design of \imis is a non-blocking traffic processing pipeline. As illustrated in Figure~\ref{fig:overview}, architecturally, \imis is designed around stateful, single-threaded tasks, which we call \emph{engines}. 
The parser engine uses DPDK~\cite{dpdk} (version 20.11) APIs to consistently collect the packet bytes from the escalated traffic; the pool engine takes the stream data as input and organizes it into per-flow state; the analyzer engine calls the pool engine to collect a batch of fresh per-flow data, and uses CUDA (version 11.7)~\cite{cuda} to interact with an auxiliary GPU to accelerate model inference; and the buffer engine stops packets without inference results to wait in memory, and sends those who have inference results to NIC.
The pool engine is the key to dynamically coordinate the speeds of the parser engine and analyzer engine, thus achieving a non-blocking packet processing pipeline. The detailed architecture of the \imis system is deferred to \S~\ref{app:subsubsec:imis}. 







\vspace{-0.2cm}
\section{Evaluation}
\vspace{-0.2cm}

\subsection{Experiment Setup}
\label{subsec:experiment_setup}

\parab{Testbed Setup.} 
We deploy our binary RNN model using one pipe of a Barefoot \mbox{Tofino 1} programmable switch. 
One server generates network traffic to an inbound port of the switch based on the pcap files we created for various traffic analysis tasks and traffic loads. 
Each flow is either analyzed by the on-switch RNN or redirected from one specific switch port to an off-switch server that deploys \imis. 
For the flows analyzed on-switch, we develop a dedicated on-switch module to collect their analysis statistics online. 
Scaling the on-switch analysis of \sys beyond a single pipe of the switch is feasible given proper flow management. We discuss this in \S~\ref{appendix:testbed}. 





\parab{Tasks.} We evaluate \sys using the following four tasks. 
\first Encrypted traffic classification on VPN: this task classifies network traffic encrypted by VPNs. We use the ISCXVPN2016~\cite{draper2016characterization} dataset, a six-class classification task (Email, Chat, Streaming, FTP, VoIP, P2P). 
\second Botnet traffic classification: this task classifies  botnet traffic collected from the IoT systems. We use the BOTIOT~\cite{koroniotis2019towards} dataset, a four-class classification task (Data Exfiltration, Key Logging, OS Scan, Service Scan).
\third Behavioral analysis of IoT devices: this task classifies traffic generated by IoT devices in different working states. We use the CICIOT2022~\cite{dadkhah2022towards} dataset, a three-class classification task (Power, Idle, Interact).
\fourth P2P application fingerprinting: this task classifies network traffic generated by P2P applications. We use the PeerRush dataset~\cite{rahbarinia2013peerrush}, a three-class classification task (eMule, uTorrent, and Vuze). 
We supplement additional details regarding the processing of these datasets in  \S~\ref{app:datasets}. 

\parab{Network Load.} 
We would like to evaluate \sys under different network loads. Similar to prior art~\cite{netbeacon,N3IC,xing2020netwarden}, we use the number of new flows arrived in each second to represent the network load. 
Specifically, for each testing flow in a task, we extract its raw packets from the pcap file while preserving the inter-packet delays. 
Given the total number of flows in this task, and a desired network load, we calculate the total time period required to replay these flows, and then uniformly release these flows within this period. If the period is too short, we replay these flows multiple times in a loop to create consistent loads throughout our test.  

The actual network load varies in different deployment. We make load estimations based on prior measurements. 
In 2015, Meta~\cite{roy2015inside} reported that its external-facing web server generates 500 new flows per second (median). Meanwhile, CISCO~\cite{CISCO} measures that Internet traffic grows 3-fold from 2016 to 2021. Combining these measurements, we estimate that 2000 new flows per second are a reasonable network load that \sys may face in practice. 
In our scaling test (see \S~\ref{subsec:deepdive}), we stress test \sys with up to 450,000 new flows per second on our testbed (a 225x increase from the normal load, and 30-300x over NetBeacon~\cite{netbeacon}). 


\parab{Metrics.} We use packet-level macro-F1 (the average of F1-score for different classes) as the accuracy metric, and further report the breakdown of the Precision / Recall of each class.


\begin{table}[t]
\centering
\caption{Experimental settings.}
\label{table:setup}
\resizebox{\linewidth}{!}{
\begin{threeparttable}
\begin{tabular}{ccccccccccccc}
    \toprule
    
    \textbf{\makecell[c]{Datasets\\(Tasks)}} & \multicolumn{3}{c}{\makecell[c]{ISCXVPN\\2016}}  & \multicolumn{3}{c}{\makecell[c]{BOT\\IOT}} & \multicolumn{3}{c}{\makecell[c]{CICIOT\\2022}} & \multicolumn{3}{c}{\makecell[c]{Peer\\Rush}}\\
    
    \midrule

    Training Flows & \multicolumn{3}{c}{7801} & \multicolumn{3}{c}{7835} & \multicolumn{3}{c}{5332} & \multicolumn{3}{c}{30770} \\

    Testing Flows & \multicolumn{3}{c}{1951} & \multicolumn{3}{c}{1961} & \multicolumn{3}{c}{1335} & \multicolumn{3}{c}{7694} \\

    Classes & \multicolumn{3}{c}{6} & \multicolumn{3}{c}{4} & \multicolumn{3}{c}{3} & \multicolumn{3}{c}{3} \\
    
    Class Ratio$\star$ & \multicolumn{3}{c}{2:6:1:5:9:3} & \multicolumn{3}{c}{1:1:4:19} & \multicolumn{3}{c}{1:4:1} & \multicolumn{3}{c}{2:1:1} \\   
    
    Best Loss & \multicolumn{3}{c}{$\mathcal{L}_{1}$} & \multicolumn{3}{c}{$\mathcal{L}_{1}$} & \multicolumn{3}{c}{$\mathcal{L}_{2}$} & \multicolumn{3}{c}{$\mathcal{L}_{1}$} \\

    $\lambda, \gamma$ & \multicolumn{3}{c}{0.8, 0} & \multicolumn{3}{c}{0.5, 0.5} & \multicolumn{3}{c}{3, 1} & \multicolumn{3}{c}{1, 0} \\

    Optimizer & \multicolumn{3}{c}{AdamW} & \multicolumn{3}{c}{AdamW} & \multicolumn{3}{c}{AdamW} & \multicolumn{3}{c}{AdamW} \\

    Learning Rate & \multicolumn{3}{c}{0.01} & \multicolumn{3}{c}{0.005} & \multicolumn{3}{c}{0.005} & \multicolumn{3}{c}{0.005} \\

    RNN Hidden States$\dag$ & \multicolumn{3}{c}{9 bits} & \multicolumn{3}{c}{8 bits} & \multicolumn{3}{c}{6 bits} & \multicolumn{3}{c}{5 bits} \\

    Per-packet Model Acc. & \multicolumn{3}{c}{0.596} & \multicolumn{3}{c}{0.327} & \multicolumn{3}{c}{0.759} & \multicolumn{3}{c}{0.684} \\
    
    \midrule
    
    \textbf{Network Load} & \multicolumn{3}{c}{Low} & \multicolumn{3}{c}{Normal} & \multicolumn{3}{c}{High} & \multicolumn{3}{c}{Scaling} \\

    \midrule

    No. of flows / s & \multicolumn{3}{c}{1000} & \multicolumn{3}{c}{2000} & \multicolumn{3}{c}{4000} & \multicolumn{3}{c}{up to 7.8M} \\
    
    \bottomrule
\end{tabular}

\begin{tablenotes}
    \item[$\star$] See \S~\ref{app:datasets} for the accurate numbers of flows in each class. 
    \item[$\dag$] We evaluate \sys under different binary RNN model sizes in \S~\ref{app:complexity}. 
\end{tablenotes}

\end{threeparttable}
}
\end{table}

\begin{table*}[t]
\centering
\caption{Analysis accuracy for \sys and other two closely related art.}
\label{table:end2end}
\resizebox{\linewidth}{!}{
\begin{threeparttable}
\begin{tabular}{c|ccc|ccc|ccc}
    \toprule
    
    Methods & \multicolumn{3}{c|}{\sys} & \multicolumn{3}{c|}{NetBeacon~\cite{netbeacon} (Tree-based Models)} & \multicolumn{3}{c}{N3IC~\cite{N3IC} (Binary MLP)} \\
    
    \midrule
    
    Network Load & Low & Normal & High & Low & Normal & High
    & Low & Normal & High \\

    \midrule
    
    \multicolumn{10}{c}{Encrypted Traffic Classification on VPN (ISCXVPN2016)} \\
    
    \midrule
    
    Email & 0.935 / 0.933 & 0.936 / 0.925 & 0.933 / 0.923 & 0.309 / 0.514 & 0.315 / 0.524 & 0.320 / 0.525 & 0.347 / 0.326 & 0.354 / 0.339 & 0.367 / 0.350 \\

    Chat & 0.903 / 0.818 & 0.902 / 0.818 & 0.901 / 0.814 & 0.739 / 0.935 & 0.739 / 0.933 & 0.742 / 0.925 & 0.336 / 0.655 & 0.336 / 0.654 & 0.342 / 0.656 \\

    Streaming & 0.926 / 0.941 & 0.926 / 0.939 & 0.926 / 0.910 & 0.963 / 0.919 & 0.962 / 0.904 & 0.962 / 0.874 &  0.741 / 0.608 & 0.742 / 0.603 & 0.743 / 0.581 \\
    
    FTP & 0.973 / 0.928 & 0.973 / 0.926 & 0.973 / 0.922 & 0.946 / 0.659 & 0.946 / 0.655 & 0.947 / 0.654 & 0.563 / 0.396 & 0.567 / 0.396 & 0.575 / 0.397 \\
    
    VoIP & 0.968 / 0.958 & 0.968 / 0.958 & 0.968 / 0.957 & 0.938 / 0.882 & 0.939 / 0.881 & 0.939 / 0.882 & 0.883 / 0.783 & 0.884 / 0.782 & 0.886 / 0.787 \\

    P2P & 0.905 / 0.927 & 0.903 / 0.928 & 0.876 / 0.930 & 0.810 / 0.959 & 0.798 / 0.959 & 0.778 / 0.960 & 0.578 / 0.739 & 0.577 / 0.742 & 0.565 / 0.748 \\

    Macro-F1 & 0.926 & 0.925 & 0.919 & 0.786 & 0.784 & 0.780 & 0.565 & 0.567 & 0.568 \\

    \midrule

    \multicolumn{10}{c}{Botnet Traffic Classification on IoT (BOTIOT)} \\
    
    \midrule

    Data Exfiltration & 0.964 / 0.974 & 0.951 / 0.973 & 0.899 / 0.971 & 0.691 / 0.845 & 0.684 / 0.847 & 0.658 / 0.848 & 0.514 / 0.879 & 0.508 / 0.881 & 0.506 / 0.879 \\

    Key Logging & 0.960 / 0.946 & 0.961 / 0.962 & 0.959 / 0.902 & 0.921 / 0.425 & 0.921 / 0.419 & 0.918 / 0.399 & 0.055 / 0.033 & 0.058 / 0.033 & 0.052 / 0.031 \\

    OS Scan & 0.996 / 0.996 & 0.995 / 0.989 & 0.995 / 0.966 & 0.838 / 0.963 & 0.841 / 0.963 & 0.844 / 0.945 & 0.831 / 0.693 & 0.830 / 0.677 & 0.831 / 0.672 \\

    Service Scan & 0.993 / 0.992 & 0.986 / 0.973 & 0.979 / 0.978 & 0.928 / 0.876 & 0.927 / 0.870 & 0.917 / 0.858 & 0.845 / 0.663 & 0.830 / 0.664 & 0.840 / 0.663 \\

    Macro-F1 & 0.978 & 0.974 & 0.955 & 0.785 & 0.782 & 0.769 & 0.547 & 0.542 & 0.541 \\

    \midrule

    \multicolumn{10}{c}{Behavioral Analysis of IoT Devices (CICIOT2022)} \\
    
    \midrule

    Power & 0.926 / 0.887 & 0.924 / 0.882 & 0.921 / 0.882 & 0.819 / 0.726 & 0.820 / 0.724 & 0.817 / 0.724 & 0.639 / 0.750 & 0.640 / 0.750 & 0.640 / 0.748 \\

    Idle & 0.922 / 0.943 & 0.921 / 0.942 & 0.918 / 0.941 & 0.810 / 0.938 & 0.808 / 0.938 & 0.806 / 0.936 & 0.618 / 0.640 & 0.620 / 0.642 & 0.622 / 0.646 \\ 

    Interact & 0.934 / 0.946 & 0.934 / 0.948 & 0.934 / 0.943 & 0.871 / 0.786 & 0.873 / 0.786 & 0.872 / 0.784 & 0.651 / 0.504 & 0.655 / 0.506 & 0.661 / 0.510 \\ 

    Macro-F1 & 0.926 & 0.925 & 0.923 & 0.822 & 0.821 & 0.820 & 0.629 & 0.631 & 0.633 \\

    \midrule
    
    \multicolumn{10}{c}{P2P Application Fingerprinting (PeerRush)} \\
    
    \midrule

    eMule & 0.943 / 0.949 & 0.918 / 0.949 & 0.898 / 0.950 & 0.846 / 0.954 & 0.821 / 0.955 & 0.805 / 0.954 & 0.734 / 0.866 & 0.730 / 0.867 & 0.723 / 0.875 \\

    uTorrent & 0.949 / 0.924 & 0.950 / 0.912 & 0.941 / 0.894 & 0.882 / 0.870 & 0.885 / 0.858 & 0.885 / 0.831 & 0.734 / 0.789 & 0.735 / 0.790 & 0.738 / 0.783 \\

    Vuze & 0.946 / 0.962 & 0.945 / 0.947 & 0.941 / 0.930 & 0.910 / 0.810 & 0.907 / 0.790 & 0.904 / 0.793 & 0.821 / 0.626 & 0.826 / 0.622 & 0.826 / 0.616 \\

    Macro-F1 & 0.945 & 0.937 & 0.925 & 0.877 & 0.866 & 0.858 & 0.755 & 0.755 & 0.752 \\
    
    \bottomrule
\end{tabular}

\end{threeparttable}
}

\end{table*}

\vspace{-0.2cm}
\subsection{End-to-end Performance}\label{subsec:end-to-end}
\vspace{-0.2cm}
In this section, we report the end-to-end performance of \sys for different tasks. The main experimental settings are summarized in Table~\ref{table:setup}. 
In \S~\ref{subsec:deepdive}, we evaluate \sys under a variety of settings. We also compare \sys with two recent art NetBeacon~\cite{netbeacon} and N3IC~\cite{N3IC}. N3IC deploys the binary MLP on a SmartNIC. For fair comparison, we simulate the switch-side traffic management logic and the binary MLP inference in software to obtain the traffic analysis results for N3IC. 
The detailed descriptions about the reproduced versions of the two art are given in \S~\ref{app:reproduce}.



\parab{Accuracy.}
We summarize the analysis accuracy results in Table~\ref{table:end2end}. 
Across all evaluated tasks, \sys achieves significantly better performance than NetBeacon and N3IC, with an average F1-score improvement of 0.13 and 0.31, respectively. On more challenging tasks with more classification classes, the improvement is even greater, up to 0.19 and 0.42, respectively.
We observe the binary MLP performs the worst because the accuracy loss caused by binarizing all model weights is significant. In fact, on the ISCXVPN2016 and CICIOT2022 datasets, the F1-scores of N3IC are even lower than these of our fallback tree-based model (0.596/0.759).
Constrained by the availability of flow features, NetBeacon can only execute model inference at discrete locations. Thus, an inference error affects all its subsequent packets until it is corrected by the next inference point. This fundamentally limits its F1-scores, especially for more difficult tasks. 
In contrast, \sys retains full-precision model weights in the on-switch RNN model and continuously produces fresh inference results as a flow proceeds. Together with the co-located \imis, \sys produces more accurate analysis results than existing arts, achieving over 0.920 F1-score in all tasks. 
We observe very minor declines of F1-scores in \sys as the network load increases, demonstrating the effectiveness of our flow management (note that we use the same flow management module for other two systems as well). The minor accuracy loss is because a small fraction of flows (\eg 2.77\%/1.43\%/2.05\%/5.22\% in the normal network load case) fall back to using the per-packet model. 


\begin{table}[t]
\centering
\caption{Hardware resource utilization.}
\label{table:storage}
\resizebox{\linewidth}{!}{
\begin{threeparttable}
\begin{tabular}{c|c|cccc}
    \toprule
    
    \multicolumn{2}{c}{\textbf{\makecell[c]{Datasets\\(Tasks)}}} & \multicolumn{1}{c}{\makecell[c]{ISCXVPN\\2016}}  & \multicolumn{1}{c}{\makecell[c]{BOT\\IOT}} & \multicolumn{1}{c}{\makecell[c]{CICIOT\\2022}} & \multicolumn{1}{c}{\makecell[c]{Peer\\Rush}}\\

    \midrule

    \multirow{6}{*} {SRAM} & \textsf{Flow Info.} (stateful) & 5.21\% & 5.21\% & 5.21\% & 5.21\% \\

    & \textsf{EV} (stateful) & 3.65\% & 3.65\% & 3.65\% & 3.65\% \\

    & \textsf{CPR} (stateful) & 5.63\% & 3.75\% & 2.81\% & 2.81\% \\

    & \textsf{FE} (stateless) & 2.19\% & 2.19\% & 2.19\% & 2.19\% \\

    & \textsf{GRU} (stateless) & 3.02\% & 1.56\% & 0.73\% & 0.73\% \\

    & Total$\star$ & 23.44\% & 20.10\% & 18.33\% & 18.33\% \\
    
    \midrule

    TCAM & \textsf{Argmax} (Total) & 1.74\% & 1.04\% & 0.69\% & 0.69\% \\ 
    
    \bottomrule
\end{tabular}

\begin{tablenotes}
    \item[$\star$] Including other components not listed, \eg packet counters for each flow.
\end{tablenotes}

\end{threeparttable}
}
\end{table}

\parab{Hardware Resource Utilization.}
We report the stateful SRAM and stateless SRAM/TCAM usage by \sys on the programmable switch in Table~\ref{table:storage}. The stateful SRAMs are consumed to maintain per-flow states, which mainly consist of the flow management information (\eg $\textsf{TrueID}$ and $\textsf{timestamp}$, see  \S~\ref{app:flow_management}), the embedding vectors (\textsf{EV}) for binary RNN inference, and the cumulative probability counter (\textsf{CPR}) for each class. In our prototype, the hardware consumption for the first two parts is task-irrelevant, and one task uses roughly 8.85\% of SRAM. The consumption for the last part depends on the number of classification classes in a task, and the four tasks use roughly 5.63\%/3.75\%/2.81\%/2.81\% of SRAM, respectively. 
The embedding vectors stored for each flow take $8\times (S{-}1) + 8$ bits (64 bits in our prototype). 
Compared with existing approaches~\cite{netbeacon,N3IC} that require online feature computation, their per-flow storage consumption depends on the used flow features. For instance, NetBeacon~\cite{netbeacon} engineers 7 important features for the P2P application fingerprinting task, which consumes roughly 150 bits.

The stateless SRAM is used to implement the lookup tables for feature embedding (\textsf{FE}) and \textsf{GRU} layers in our binary RNN. Specifically, the SRAM consumption of \textsf{GRU} layers depends on the number of bits used for storing RNN hidden states. Using the default bitwidth in Table~\ref{table:setup}, the four tasks use roughly 3.02\%/1.56\%/0.73\%/0.73\% of SRAM, respectively, for \textsf{GRU} layers. Additionally, each task uses 2.19\% of SRAM for feature embedding. In total, the four tasks use 23.44\%/20.10\%/18.33\%/18.33\% of SRAM, respectively. 

\sys uses TCAM to implement the \textsf{argmax} operation. 
Compared with NetBeacon~\cite{netbeacon}, \sys consumes SRAM of similar size and 20x less TCAM (note that the ternary matching in TCAM is 6x more expensive than exact matching with SRAM, in terms of required silicon resources~\cite{N3IC}).

\begin{figure*}[t]
    \centering
    \includegraphics[width=\textwidth]{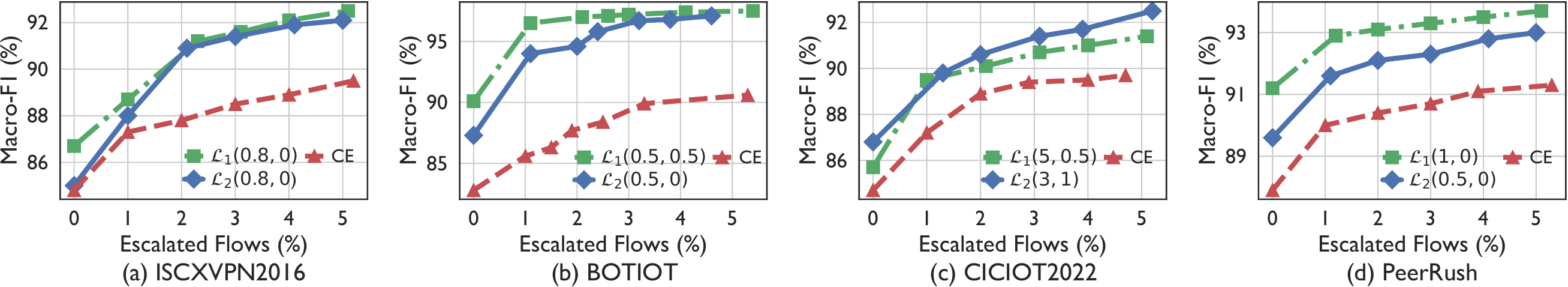}
    \caption{\label{fig:tradeoff} [Testbed] The trade-off between percentage of escalated flows and the overall accuracy.}
\end{figure*}

\begin{figure*}[t]
    \centering
    \includegraphics[width=\linewidth]{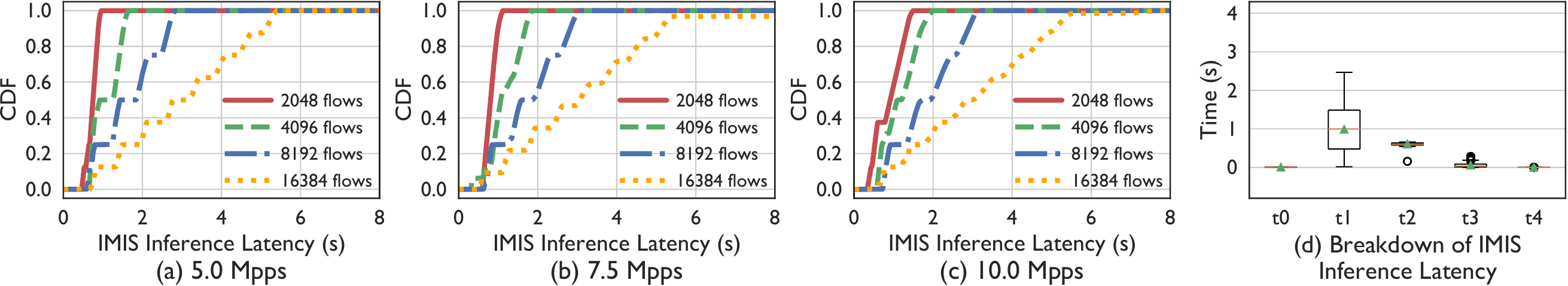}
    \caption{\label{fig:scalability_imis} [Testbed] The inference throughput and latency of the off-switch \imis.}
\end{figure*}


\begin{figure*}[t]
    \centering
    \includegraphics[width=\linewidth]{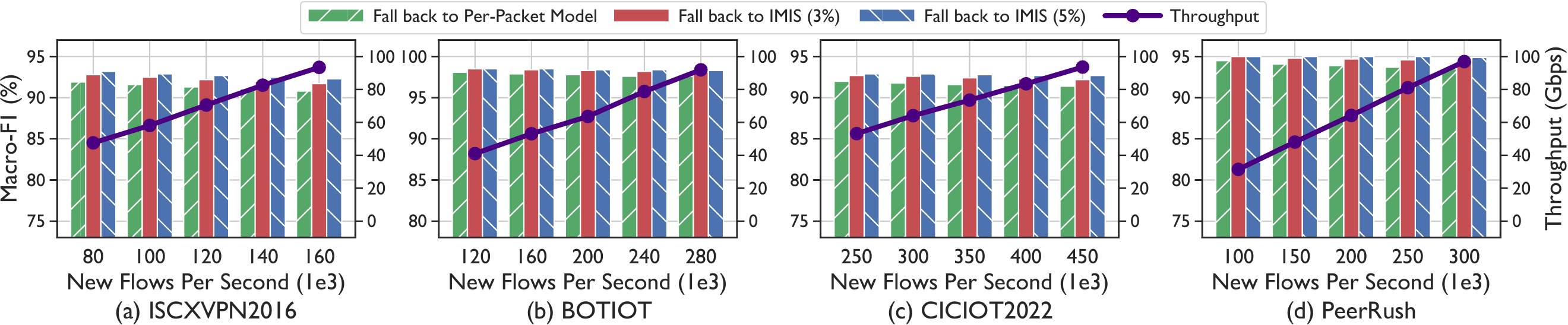}
    \caption{\label{fig:scalability_sys_real} [Testbed] Scaling test of \sys when we progressively increase the aggregate throughput to 100 Gbps.}
\end{figure*}

\begin{figure*}[t]
    \centering
    \includegraphics[width=\linewidth]{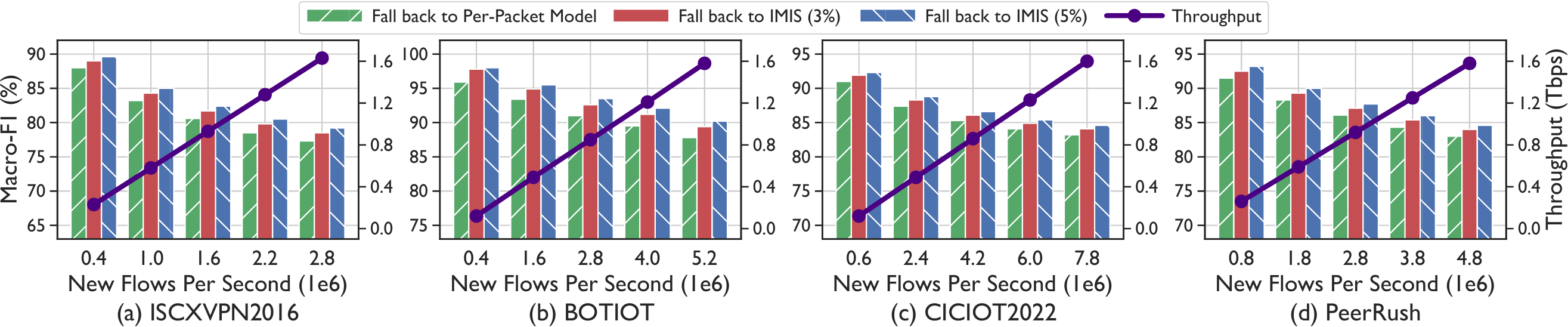}
    \caption{\label{fig:scalability_sys_simulation} [Simulation] 
Scaling test of \sys when we progressively increase the aggregate throughput to 1.6 Tbps.}
\end{figure*}


\subsection{\sys Deep Dive}
\label{subsec:deepdive}

\parab{Analysis Escalation.} In this segment, we study the trade-off between the amount of escalated flows and the overall macro-F1, demonstrating that our loss functions defined in \S~\ref{subsec:ana_escalation}
achieve a better trade-off than the classic cross entropy loss.  
As described in \S~\ref{subsec:ana_escalation}, the escalation threshold $T_\textsf{esc}$ controls the amount of escalated flows. Using the setting in Table~\ref{table:setup}, we train the binary RNN with our losses and cross entropy loss, respectively, and measure the overall macro-F1 with different amount of escalated flows under the normal network load (2000 flows/s). The results are plotted in Figure~\ref{fig:tradeoff}, and the best parameters ($\lambda, \gamma$) of our losses in each task are presented. We make the following key observations. \first Regardless of the used loss functions, the overall macro-F1 scores for all tasks improve as the percentage of escalated flows increases from 0\% to 5\%. This demonstrates the necessity for accommodating the off-switch analysis model to compensate for on-switch analysis. 
\second For the same amount of escalated flows, our losses outperform the cross entropy loss by significant margins across all tasks. This shows that our loss designs can more effectively identify the ambiguous packets that require additional reevaluation. This is crucial to improve the overall system performance without consistently escalating flows.  
\third The performance of our two losses $\mathcal{L}_{1}$ and $\mathcal{L}_{2}$ is task-dependent. In general, $\mathcal{L}_{1}$ outperforms $\mathcal{L}_{2}$ in three tasks, yet $\mathcal{L}_{2}$ requires less training epochs to converge.

\parab{System Performance of \imis.}
In this segment, we stress test the performance of off-switch \imis upon a burst of concurrent flows.  We run the \imis with 8 parallel analysis modules. 
We evaluate four different levels of flow concurrency (2048, 4096, 8192, and 16384 flows) with three different aggregate inbound rates (5.0, 7.5 and 10.0 million packets per second). The complete inference pipeline for a packet $\mathcal{P}$ in \imis has six phases: (1) $\mathcal{P}$ is  fetched from the NIC by the parser engine; (2) its metadata is organized by the pool engine; (3) its metadata is sent to the analyzer engine; (4) the analyzer engine generates the inference result; (5) the result is collected by the buffer engine; (6) $\mathcal{P}$ is dispatched to NIC by the buffer engine. 

The transformer model in \imis performs inference on the first five packets of each flow. Given that the average length of the escalated flows in each tasks is 801, 255, 167, and 138 packets, respectively, the vast majority of packets in these escalated flows 
are directly forwarded to the buffer engine after being collected from the NIC, experiencing very minor latency (less than 1ms). In the following, we only consider the latency for the packets that traverse the entire inference pipeline.
The CDFs of the end-to-end latencies are plotted in Figures \ref{fig:scalability_imis}(a) to (c). When the number of concurrent flows is below 4096, the maximum end-to-end latency imposed by \imis is less than 2 seconds even for \mbox{10.0 Mpps} inbound rate (equivalently \mbox{41 Gbps} as the packet sizes we send are \mbox{512 B}). 
Considering that \sys typically escalates less than 5\% of flows, the flow concurrency levels experienced by the \imis are expected to be low in most deployments. 
In Figure \ref{fig:scalability_imis}(d), we further report the breakdown of the end-to-end latency (\ie the time intervals between two consecutive phases in the inference pipeline) under 8192 concurrent flows and an inbound rate of \mbox{5.0 Mpps}. We observe that the major latency occurs between the second and third phase, when the packets are waiting to be collected by the analyzer engine. The net inference time spent in the analyzer engine is about \mbox{0.6 s}.

\parab{Scaling Test.} 
We stress test \sys in high-throughput scenarios with high flow concurrency and high flow throughput. 
Specifically, because all the original network traces are collected in low bandwidth networks (\eg tens of Mbps), we create high-throughput network traces by concurrently packaging a large number of flows (while ensuring each flow has a unique identifier) and accelerating the packet replay speeds (by reducing the inter-packet delays). Then we replay these pcap files to generate traffic on our testbed. 
Figure~\ref{fig:scalability_sys_real} presents the scaling test results, 
where we progressively increase the flow concurrency to saturate the physical capacity of the NIC on our traffic generator. 
The results demonstrate that \sys can comfortably handle this level of scale, as the macro-F1 scores remain nearly identical compared to the results in Table~\ref{table:end2end}. 

To evaluate \sys at even larger scales, we build a simulator to emulate the entire workflow of \sys. 
The accuracy of the simulator is validated by replicating the experimental settings of Table~\ref{table:end2end} and Figure~\ref{fig:scalability_sys_real}. 
The accuracy results obtained through the simulation are almost the same as those collected from our testbed. 
We subsequently employ the simulator to explore significantly larger scales, 
progressively increasing flow concurrency to up to 7.8 million flows per second, and the aggregate throughput to over \mbox{1.6 Tbps}. 
The results depicted in Figure~\ref{fig:scalability_sys_simulation} reveal a sublinear decline in the macro-F1 scores of \sys, culminating in a ${\sim}11.6$\% reduction at the largest scale.

\parab{Fallback Alternative.}
The default handling of the flows without dedicated per-flow storage is to analyze their packets using a tree-model trained trained only on per-packet features (see \S~\ref{app:perpacket}). Alternatively, a subset of the flows without dedicated per-flow storage can be forwarded to a new instance of off-switch \imis dedicated to handling these flows. In Figure~\ref{fig:scalability_sys_real} and~\ref{fig:scalability_sys_simulation}, we report the macro-F1 when forwarding a certain percentage of flows without per-flow storage to a dedicated \imis. When the flow concurrency is high (\ie Figure~\ref{fig:scalability_sys_simulation}), this method exhibits reasonable accuracy advantages over falling back to use the per-packet model. 

\vspace{-0.2cm}
\section{Discussion and Related Work}
\vspace{-0.2cm}

\parab{Hardware Dependency.}
\sys is generic in the sense that all its core designs (\eg retaining full-precision RNN model weights, using sliding windows to compute unlimited RNN step times) are all realizable using match tables. 
Since table matching is the universal primitive for any data plane, we expect our designs, with lightweight adaptation, are also deployable on other types of programmable data plane devices. 

\parab{ML-driven Traffic Analysis.} 
Our community has proposed various ML-powered traffic analysis designs, such as intrusion detection~\cite{mirsky2018kitsune, jan2020throwing, fu2021realtime}, website fingerprinting~\cite{shen2020fine, yin2021automated, rimmer2018automated, deng2023robust}, and encrypted traffic classification~\cite{van2020flowprint, shen2021accurate, siby2020encrypted, qing2024low}. 
However, it is difficult to directly apply their models in \idp due to the hardware constraints on the data plane.  

\parab{Advances in the Programmable Data Plane.} The flexibility of programmable switches encourages a number of customized applications on the data plane, including network telemetry and monitoring~\cite{sengupta2022rtt,2022sketchlib,molero2022fast}, network security~\cite{netbeacon,xing2022bedrock,xing2021ripple,zhao2024effective}, and network functions~\cite{jung2022acl}. Additionally, \cite{lao2021atp, sapio2021scaling} use programmable switches to accelerate ML training, 
and \cite{li2019accelerating, swamy2022taurus, zhong2021ioi} design auxiliary modules within a switch or leverage off-switch FPGA. Our work focuses on enabling NN-driven \idp using only commodity hardware.

\parab{Deployment.} 
\sys is an application-specific system designed for high-throughput and low-latency NN-driven traffic analysis. Therefore, we have not discussed co-deploying other networking functions with \sys on the same programmable switch. Although \sys consumes multiple stages, the SRAM/TCAM consumption per stage is small (see Table~\ref{table:storage}). Thus, networking functions orthogonal to \sys (\eg the ECMP in~\cite{switchp4}) can be co-deployed with \sys in parallel. Additionally, the latest Tofino chips have almost doubled the number of stages and TCAM/SRAM resources compared to the \mbox{Tofino 1} chip we use. Thus, we envision that networking functions that may depend on \sys's analysis results (\eg the traffic policing in~\cite{switchp4}) may also be co-deployed with \sys.
\vspace{-0.2cm}
\section{Conclusion}
\vspace{-0.2cm}

In this paper, we present \sys, the first \idp design that enables NN-driven traffic analysis at line-speed. The key novelty of \sys is to realize complex RNN computations using a set of novel data plane native operations, and meanwhile to accommodate a transformer-based traffic analysis module via a carefully designed flow escalation mechanism. We implement a prototype of \sys and evaluate it thoroughly on four traffic analysis tasks. The results demonstrate that \sys advances SOTA in both traffic analysis accuracy and scalability.   

\vspace{-0.2cm}
\section*{Acknowledgement}   
\vspace{-0.2cm}
We thank our shepherd Costin Raiciu and the anonymous NSDI reviewers for their insightful feedback. The research is supported in part by the National Key R\&D Program of China under Grant 2022YFB2403900, NSFC under Grant 62132011 and Grant 61825204, and Beijing Outstanding Young Scientist Program under Grant BJJWZYJH01201910003011. 
The corresponding author of this paper is Zhuotao Liu. 

\balance 
\bibliographystyle{plain}
\bibliography{reference}

\clearpage
\appendix
\section{Appendix}
\subsection{Model Realization on the Data Plane}
In this segment, we supplement additional design details regarding deploying our binary RNN on the programmable data plane. 

\subsubsection{Intuitive Number Comparison}
\label{subsec:number_comparison}
The foundation to realize \textsf{argmax} is number comparison. 
There are two types of approaches to compare numbers on the programmable data plane. The first is based on conditional statements. Specifically, a simple statement to compare num$_A$ and num$_B$ is 
\begin{verbatim}
    if (num_A > num_B) act_A; else act_B;
\end{verbatim}
which, unfortunately, is not compiled because the condition is too complex. We could avoid complex condition by the following statement 
\begin{verbatim}
    tmp = num_A - num_B;
    if (tmp > 0) act_A; else act_B;
\end{verbatim}
which takes at least two switch stages. More crucially, it is difficult to scale statement-based approaches to compare $n$ numbers since it would take $n\times (n-1)/2$ differences.  

The alternative approach is designing a match-action table which takes the num$_A$ and num$_B$ as keys (\eg the concatenation of their bits) and performs act$_A$ or act$_B$ according to the lookup result. Meanwhile, we can easily extend width of keys to realize the comparison of multiple numbers, which is exactly the \textsf{argmax} operation. Yet, the drawback is the explosion of required table entries: to obtain the maximum number among $n$ values, each with $m$-bits, it takes $2^{nm}$ entries to enumerate all possible key-value combinations. 

\subsubsection{Ternary-Matching Based Design}
\label{appendix:argmax}
In \S~\ref{subsubsec:decision_making}, we discuss the implementation of argmax operation with ternary matching. 
We define the problem as following conditions and restrictions.

\begin{enumerate}
\item There are $n$ numbers of $m$ bit(s) each, where $n,m\ge 1$.
\item There is a predefined order to determine which number to select when there is a tie for maximum value.
\item Only one ternary matching is allowed, i.e., the calculation before matching is prohibited.
\end{enumerate}

We denote the number of table entries as $F(n,m)$. Based on the basic optimization in \S~\ref{subsubsec:decision_making}, we get the following recurrence relationship of $F(n,m)$.

\begin{equation}
F(n,m)=2\times F(n,m-1)+\sum_{i=1}^{n-1}\binom{n}{i}F(i,m-1),n\ge 2,m\ge 2
\end{equation}

We explain the meaning of the above equation as follows. 
The entries should cover all  possible combinations of the $n$ numbers. We consider all the combinations in different categories classified according to the combination of most significant bits (MSBs). Among all $2^n$ categories, there are $\binom{n}{i}$ categories where $i$ numbers are with $\textsf{MSB}=1$ and $n-i$ numbers are with $\textsf{MSB}=0$ ($i\in [1,n-1]$). In these categories, we do not further consider those $n-i$ numbers with $\textsf{MSB}=0$, and only focus on the possible combinations of the $i$ numbers with $\textsf{MSB}=1$, which are sub-problems with $n^{\prime}=i,m^{\prime}=m-1$ and require $\binom{n}{i}F(i,m-1)$ entries in total. In the other two categories (all $\textsf{MSB}=0$ or all $\textsf{MSB}=1$), we continue to focus on the possible combinations of all the $n$ numbers, which are both sub-problems with $n^{\prime}=n,m^{\prime}=m-1$ and require $2\times F(n,m-1)$ entries in total.

After the two optimizations described in \S~\ref{subsubsec:decision_making}, the recurrence relation of $F(n,m)$ is given as follows.

\begin{equation}
\label{math:recursion}
F(n,m)=F(n,m-1)+\sum_{i=1}^{n-1}\binom{n}{i}F(i,m-1),n\ge 2,m\ge 2
\end{equation}

\begin{equation}
\label{math:base,m=1}
F(n,1)=n,n\ge 1
\end{equation}

\begin{equation}
\label{math:base,n=1}
F(1,m)=1,m\ge 1
\end{equation}

By solving this iterative formula, we obtain $F(n,m)=nm^{n-1}$.
We provide the derivation process from Equation~(\ref{math:recursion}) as follows. First, let $F(0,m)=0,m\ge 1$, which is 
consistent with the Equation~(\ref{math:base,m=1}). Then the Equation~(\ref{math:recursion}) can be written as: 

\begin{equation}
F(n,m)=\sum_{i=0}^{n}\binom{n}{i}F(i,m-1),n\ge 0,m\ge 2
\end{equation}

Then, we transform the formula and get:
\begin{equation}
F(n,m)=\sum_{i=0}^{n}\frac{n!}{i!(n-i)!}F(i,m-1),n\ge 0,m\ge 2
\end{equation}

\begin{equation}
\label{math:transformation}
\frac{F(n,m)}{n!}=\sum_{i=0}^{n}\frac{F(i,m-1)}{i!}\frac{1}{(n-i)!},n\ge 0,m\ge 2
\end{equation}

We denote $\frac{F(n,m)}{n!}$ as $^{m}g_n$, and $\frac{1}{n!}$ as $h_n$. Then we construct the generating function of $^{m}g$ and $h$.

\begin{equation}
\label{math:generation,g}
^{m}G(x)=\sum_{n=0} {^{m}g_{n}}x^n=\sum_{n=0}\frac{F(n,m)}{n!}x^n,m\ge 1
\end{equation}

\begin{equation}
H(x)=\sum_{n=0}h_{n}x^n=\sum_{n=0}\frac{x^n}{n!}=e^x
\end{equation}

We can obtain the recursive relations between $^{m}G$ and $^{m-1}G$ from Equation~(\ref{math:transformation}).

\begin{equation}
^{m}G=^{m-1}G\times H,m\ge 2
\end{equation}

And for $m=1$, we have

\begin{equation}
\begin{aligned}
^{1}G(x)=\sum_{n=0} {^{1}g_{n}}x^n=\sum_{n=0}\frac{F(n,1)}{n!}x^n\\
=\sum_{n=0}\frac{n}{n!}x^n=\sum_{n=1}\frac{1}{(n-1)!}x^n\\
=x\sum_{n=0}\frac{1}{n!}x^n=xe^x
\end{aligned}
\end{equation}

With the mathematical induction, we can get

\begin{equation}
^{m}G(x)=xe^{mx},m\ge 1
\end{equation}

Compared with the Equation~(\ref{math:generation,g}) and we can get

\begin{equation}
\begin{aligned}
F(n,m)=^{m}G^{(n)}(0)=(x e^{m x})^{(n)}\bigg|_{x=0}\\
=(n m^{n-1}+m^n x)e^{m x}\bigg|_{x=0}\\
=nm^{n-1},n\ge 1,m\ge 1
\end{aligned}
\end{equation}

We can verify the result using  mathematical induction. 

\begin{equation}
\begin{aligned}
F(n,m-1)+\sum_{i=1}^{n-1}\binom{n}{i}F(i,m-1)\\
=\sum_{i=1}^{n}\binom{n}{i}i(m-1)^{i-1}\\
=n\sum_{i=1}^{n}\binom{n-1}{i-1}(m-1)^{i-1}\\
=n\sum_{i=0}^{n-1}\binom{n-1}{i}(m-1)^{i}\\
=nm^{n-1}=F(n,m)
\end{aligned}
\end{equation}

In Table~\ref{tab:number_of_entries}, we list the number of entries required for different combinations of $m$ and $n$. The results demonstrate that our design, augmented by two optimizations, significantly reduces table consumption for achieving the \textsf{argmax} operation. 

\begin{table}[t]
\centering
\caption{The no. of entries required for different $m,n$.}
\label{tab:number_of_entries}
\resizebox{\linewidth}{!}{
\begin{threeparttable}
\begin{tabular}{cccccc}
    \toprule

    \textbf{No. of \textsf{Entries}} & \textbf{\textsf{Opt} 1} \boldmath{$\&$} \textbf{2} & \textbf{\textsf{Opt}\ 2 only} & \textbf{\textsf{Opt}\ 1 only} & \textbf{Base Design} & \boldmath{$2^{mn}$} \\
    
    \midrule    

    n=3,m=16 & 768 & 2949123 & 863 & 4587523 & 2.81e14 \\ 
    n=4,m=8 & 2048 & 44028 & 2788 & 76028 & 4.29e9 \\ 
    n=5,m=5 & 3125 & 10245 & 5472 & 21077 & 3.36e7 \\ 
    n=6,m=4 & 6144 & 10890 & 13438 & 26978 & 1.68e7 \\ 
    
    \bottomrule
\end{tabular}
\end{threeparttable}
}
\end{table}
\subsubsection{Packet Counters}
\label{subsubsec:counters}

Because the number of packets in a flow is unknown in advance, statically allocating a fixed width of bits for packet counters may result in buffer overflow.  
Meanwhile, as described in \S~\ref{subsubsec:embedding_vector_storage}, we need to perform modulo operations on packet count (\ie $\textsf{pktcnt} ~\%~ (S{-}1)$) when storing embedding vectors. 
Thus, packet counting in \sys is designed based on two parallel counters: the first counter increases from $1$, and stops at $S$ (the sliding window size). For the $i^\text{th}$ packet, it returns $i$ if $i<S$, otherwise it returns $S$. The second counter increases from $0$ and cycles back to $0$ after $S-2$, simulating the modulo operation. Thus, when the number of arrived packets in the flow exceeds $S$, the first counter essentially becomes a flag indicating that index for the ring buffer (storing embedding vectors) can be read from the second counter. 

\subsubsection{Flow Management}
\label{app:flow_management}

\sys relies on stateful storage to maintain per-flow state. 
Prior art~\cite{barradas2021flowlens, xing2020netwarden} relies on the control plane to allocate non-conflicting storage indices for different flows. 
To achieve line-speed traffic analysis, \sys relies on the readily available hardware hashing to allocate flow storage indices. 
In particular, the storage index for flow $f$ is computed as $\mathcal{H}(f(\textsf{5-tuple}) ~\%~ \textsf{N})$, where $\mathcal{H}$ is the hash function, and $\textsf{N}$ represents the total number of continuous per-flow storage blocks allocated for maintaining per-flow state. 

Both hash and modulo operations may result in flow index collisions, \ie two different flows (with different 5-tuples) may receive the same storage index. To avoid confusions, \sys stores a tuple $\{\textsf{TrueID}, \textsf{timestamp}\}$ alongside the storage index, where $\textsf{TrueID}$ represents the actual flow  identifier\footnote{To avoid resubmitting or recirculating packets, the read and write of the tuple need to be finished in an atomic operation. This restricts the length of $\textsf{TrueID}$ so that we cannot directly use 5 tuple as the $\textsf{TrueID}$. Thus, we leverage a different hash function $H^{\prime}$ to calculate the $\textsf{TrueID}$ as $H^{\prime}(f(\textsf{5-tuple}))$.} and \textsf{timestamp} represents the latest packet arrival time for the flow. 
When storage indices collide, \sys allows the new flow to take the occupied storage only if the existing flow is timed out (\ie the stored \textsf{timestamp} is earlier than a predefined threshold). Otherwise, the new arrived flow falls back to use the per-packet tree model trained using only per-packet features, or falls back to \imis; see discussions in \S~\ref{subsec:deepdive}. 

When developing the prototype of \sys, we observe a possible corner case for flow management. Specifically, the switch has multiple forwarding pipes, each of which has several processing stages. 
To support more complex RNN models, we can simultaneously use the stages in both the ingress and egress pipes. However, if
multiple ingress pipes could be mapped to the same egress pipe (\eg traffic entered from both pipe A and pipe B may exit from pipe A), we would need to deploy a flow management module in both the ingress and egress pipe, because flows that do not collide in their ingress pipes may collide in the egress pipe. We have not encountered this corner case even in our scaling experiments (see \S~\ref{subsec:deepdive}), and therefore we only deploy the flow management module in the ingress pipe. 

\subsubsection{Per-packet Fallback Model}
\label{app:perpacket}
When the flow manager cannot allocate storage for a new flow, \sys falls back to analyzing the packets of that flow using a tree
model trained only using per-packet features. Specifically, we use a 2$\times$9 Random Forest model (2 trees with max depth 9), and use the same per-packet features as in~\cite{netbeacon} (\eg packet length, TTL, Type of Service, TCP offset). We apply the coding mechanism from NetBeacon~\cite{netbeacon} to deploy this tree model on the data plane alongside our binary RNN model.

\subsubsection{The Pre-analysis Issue}
\label{app:preanalysis}
As discussed in \S~\ref{subsec:online-inference}, we employ a sliding window mechanism in our binary RNN inference where the model continuously processes packet segments. The length of the segment is a hyper-parameter $S$ (set to 8 in our prototype). 
As a result, the very first $S{-}1$ packets of a flow cannot form a complete segment. This results in the pre-analysis issue: the inference results on these packets may be inaccurate because the model simply has not observed enough information. Any model-driven (or data-driven) traffic analysis approach has this limitation. 

To avoid premature inference results caused by the pre-analysis problem, \sys regards the first $S{-}1$ packets of a flow as \emph{pre-analysis packets}, and only starts to produce inference results for the $S^\text{th}$ and subsequent packets (\ie any inference result output by \sys is based on at least one full segment).  
The protocol for forwarding these pre-analysis packets should be application-specific. For instance, in security-oriented task, \sys can forward pre-analysis packets via a dedicated low priority queue so that a strategic adversary cannot overwhelm the network by sending very short flows (less than $S$ packets). 
In other tasks (\eg an inbound gateway on a campus/enterprise that loads balance different types of traffic received from the Internet), simply forwarding these pre-analysis packets may be sufficient, considering the average length of campus Internet flows (${\sim}120$)~\cite{2018campus} is much larger than $S$ (8 in our prototype). 
Finally, it is possible to employ another learning model trained only on per-packet features (such as~\cite{xie2022mousika}) to process these pre-analysis packets. 


\subsection{Prototype Implementation}
\label{app:implementation}

In this segment, we supplement additional details about our implementation. 

\subsubsection{On-Switch RNN Inference}
\label{app:subsubsec:brnn}

\parab{Component Overview.}
We plot the hardware implementation of the binary RNN in \sys in Figure \ref{fig:prototype}. 
The top part shows the simplified dependency graph of all components. We plots two types of dependency: if $a$ has data dependency on $b$, then the input data of $a$ is (partially) provided by $b$; if $a$ has control dependency on $b$, then the execution of $a$ is determined by $b$. 
For instance, the GRU tables have data dependency on embedding vector storage; the window CTR (counting the number of windows/segments) has control dependency on the PKT CTR-1 (indicating whether the number of received packet is no less than $S$).

\parab{Per-Stage Breakdown.}
The bottom-right part shows the breakdown of stage usage for deploying a \sys model with the hyper-parameters shown in the bottom-left part. We use the stages in both ingress and egress pipeline. 
The $k^\text{th}$ ingress stage and $k^\text{th}$ egress stage share the same underlying hardware resource. 

We first introduce the stage usage in ingress. In the stage $0$, besides calculating the flow index and \textsf{TrueID} for flow management, it also executes embedding of packet length, since it has no other dependency. 
Then, the \textsf{FlowInfo} tuple (\ie $\{ \textsf{TrueID, timestamp} \}$) is stored in stage $1$ for flow collision avoidance. Flow management is only necessary in the ingress pipeline (see \S~\ref{app:flow_management}). 
The inter-packet delay (IPD) embedding is implemented using the following three stages: stage $2$ stores the last packet timestamp, stage $3$ obtains IPD by subtracting the current packet arrival time with the last timestamp, and stage $4$ computes the embedding of IPD. 
In stage $5$, an FC layer takes in the packet length embedding and IPD embedding to output an embedding vector, which is stored using $7$ bins and dispatched to corresponding GRU tables in stages $6$ to $8$. All seven bins cannot be allocated into one stage because only $4$ registers (register arrays) are allowed in one stage. 
In the last three stages of ingress, the first four GRU tables (\ie $\textsf{GRU}_{1,2,3,4}$) are placed sequentially.
The first two GRU tables (\ie $\textsf{GRU}_{1,2}$) are implemented with one match-action table, \ie $h{=}\textsf{GRU}_2(\textsf{GRU}_1(0,ev_1),ev_2)$ is merged as $h{=}(\textsf{GRU}_2\circ\textsf{GRU}_1)(0,ev_1,ev_2)$.

In the egress pipeline, the remaining four GRU tables (\ie GRU$_{5,6,7,8}$) are placed from stage 0 to stage 3.
The output layer is merged with GRU$_8$, \ie $C{=}\textsf{Output}(\textsf{GRU}_8(h,ev_8))$ is merged as $C{=}(\textsf{Output}\circ\textsf{GRU}_8)(h,ev_8)$. 
The counters to accumulate per-class probabilities ($\textsf{CPR}_{1..6}$) are spread in stage $4$ and $5$. 
To accumulate the probability vectors (\ie the intermediate results) on the data plane, we quantize the probability for a class to an integer from 0 to 15. Considering the reset period of 128 packets, the width of cumulative probability is $\lceil log_2 (16\times 128) \rceil{=}11$. To implement \textsf{argmax} for $n{=}6,m{=}11$, we split it into three sequential \textsf{argmax} operations, two for $n{=}3,m{=}11$ and one for $n{=}2,m{=}11$, \ie comparing the first three numbers in stage $5$, then comparing the rest three numbers in stage $6$, then comparing the two winners in stage $7$.
Finally, the escalation logic is implemented in stage $8$ to $9$. To obtain the classification confidence for a packet, we do not actually divide the largest accumulative probability with $\textsf{wincnt}$ (the total number of intermediate results for the flow), as division is not supported on the data plane. Instead, we compare the probability with $\mathbb{T}_\textsf{conf}\times\textsf{wincnt}$, which is divided into a subtraction and a comparison with $0$. 
The subtraction is performed in the \emph{action} of the winning case of \textsf{argmax} match-action table. And the comparison with $0$ is executed while reading/updating the counter that stores the number of ambiguous packets in stage $8$. If the counter exceeds $T_\textsf{esc}$, the packet is escalated to \imis.


\parab{Escalation Flag.}
Due to the limited number of stages in the ingress pipeline of the \mbox{Tofino 1} switch, we must use both the ingress and egress pipelines in our prototype. However, this poses a challenge because the egress port of a packet must be determined in the ingress pipeline, but we cannot determine whether a packet should be escalated to \imis until all operations in the egress pipeline have completed. To address this challenge, we store an escalation flag in the ingress pipeline so that the egress port for a packet can be properly determined. We update the escalation flag through egress-to-egress mirroring and recirculating.

\subsubsection{Implementation of \imis}
\label{app:subsubsec:imis}

\begin{figure}[t]
    \centering
    \includegraphics[width=0.95\linewidth]{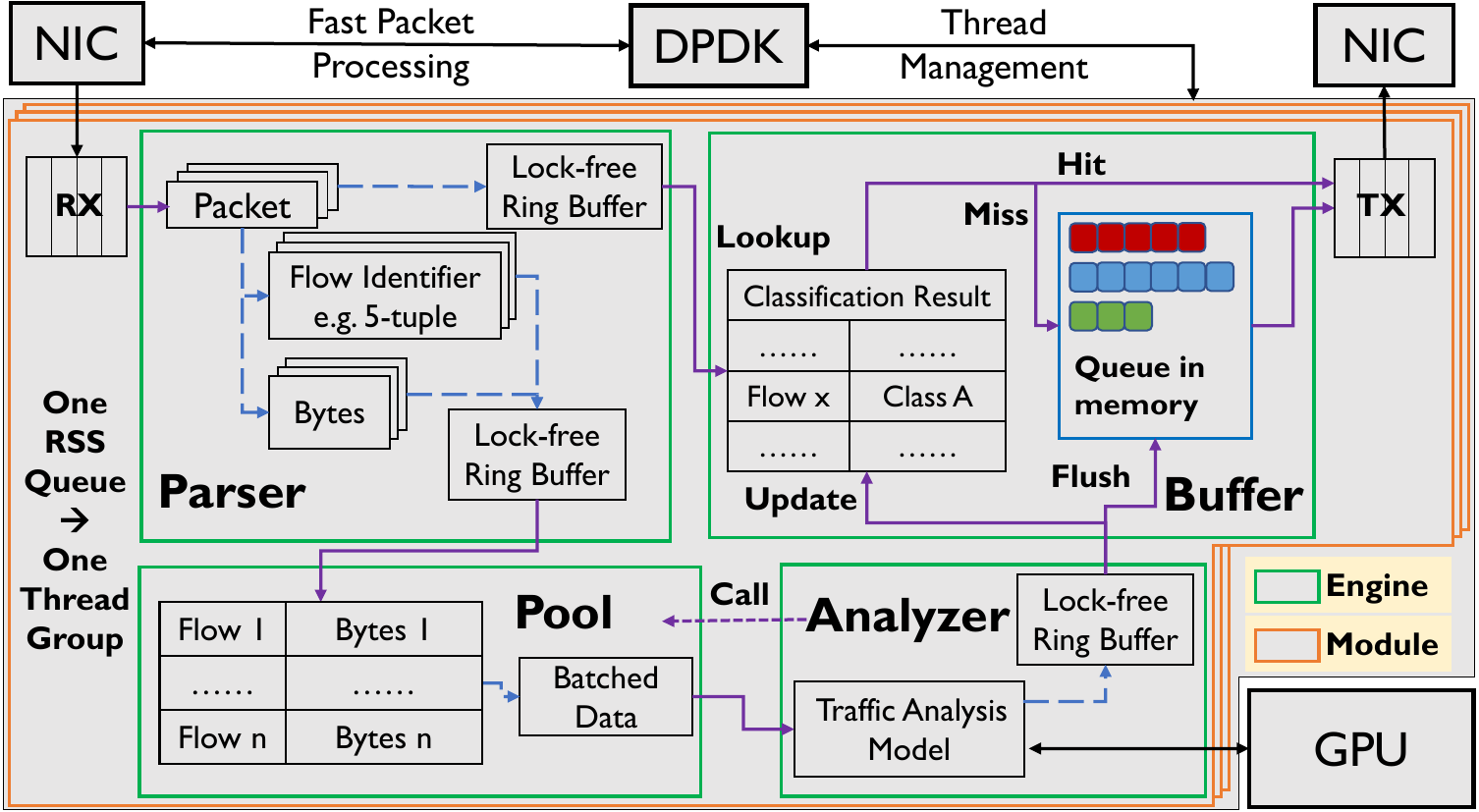}
    \caption{\label{fig:imis} The architecture of \imis.}
\end{figure}

The architecture of \imis is plotted in Figure~\ref{fig:imis}. 
We use Intel Data Plane Development Kit (DPDK version 20.11.9 LTS)~\cite{dpdk} to enable multiple NIC RX/TX queues, each of which is bound to one analysis module. The Receive Side Scaling (RSS)~\cite{rss} is enabled to efficiently distribute traffic to each analysis module.

\parab{Analysis Module.} The parser engine uses DPDK APIs to parse flow identifier (\eg 5-tuple) and the raw bytes from the input traffic. It stores the parsing results into a lock-free ring buffer consumed by the pool engine to maintain per-flow state and perform batch arrangement. After obtaining the parsing result of a packet, the packet is sent to another lock-free ring buffer consumed by the buffer engine to perform egress queuing  according to model inference results. As our transformer-based model only uses the first 5 packets in a flow for inference, the subsequent packets sent by the flow will be forwarded to the buffer engine directly without raw bytes extraction.

The pool engine translates the streamed parsing results into batched data to facilitate model inference. Specifically, it continuously fetches the raw byte features of packets from the lock-free ring buffer linked to the parser engine, and organizes them as per-flow state. 
When it receives a call from the analyzer engine, the pool engine selects flows according to their timestamps to form a batch of inputs, and sends the batch to the analyzer engine for inference. If a selected flow has fewer than 5 packets, the pool engine pads its data with zeros. The inference result obtained for this flow is considered to intermediate, and the pool engine may select this flow again in the next round. 


To accelerate model inference, the analyzer engine uses CUDA (version 11.7)~\cite{cuda} to interact with the auxiliary GPU card. Specifically, it continuously requests input batches from the pool engine. Upon receiving a batch, the analyzer engine executes inference on the GPU and sends the results to a lock-free ring buffer consumed by the buffer engine.

The buffer engine continuously fetches the latest inference results from the analyzer engine and uses the results to release packets. Upon receiving a packet from the parser engine, the buffer engine checks if the inference result for the packet's flow has been determined. If so, the packet is released immediately. Otherwise, the packet is placed in the egress queue for its flow to wait for the inference result. When the buffer engine receives a flow inference result from the analyzer engine, it releases all packets in the egress queue for that flow.

The buffer engine keeps fetching the latest inference results from the analyzer engine, and uses the results to release packets. Upon receiving a packet from the parser engine, the buffer engine checks if the inference result for the packet's flow has been determined. If so, the packet is released immediately. Otherwise, the packet is placed in the egress queue for its flow to wait for the inference result. When the buffer engine receives a flow inference result from the analyzer engine, it releases all packets in the egress queue for that flow. 


\subsection{Additional Details about Testbed}
\label{appendix:testbed}

We use a Wedge 100BF-32X programmable switch with 2 pipes and \mbox{32$\times$100 Gbps} ports to deploy the on-switch RNN in \sys. The version of SDE is 9.7.0. The off-switch \imis is hosted on server with two Intel(R) Xeon(R) Gold 6348 CPUs (2$\times$28 cores), Ubuntu 20.04.1, \mbox{512 GB} memory, and one Mellanox 100 Gbps NIC with two ports that support DPDK (version 20.11). We reserve 160 GB memory as huge pages for DPDK (\mbox{80 GB/NUMA Node}), and an NVIDIA A100 GPU is attached to \imis. All physical cores for parser engines, pool engines and buffer engines are on the same Node with NIC, and all the physical cores for analyzer engines are on the same Node with an auxiliary GPU.

\parab{Scaling the On-switch Analysis Beyond One Pipe.} The on-switch analysis of our current prototype is implemented using one switch pipe. The complexity of scaling the analysis beyond one pipe depends on whether cross-pipe traffic forwarding is allowed.  
Specifically, if the traffic forwarding for each pipe is self-contained (\ie the traffic ingressing from one pipe will only exit from this pipe), we can easily operate multiple pipes independently, where each pipe hosts an instance of our on-switch RNN and processes traffic in parallel. 
However, when the flows entering from different pipes can eventually exit via the same pipe, we have to deploy our flow management modules in both the ingress and egress pipes (as we have discussed in \S~\ref{app:flow_management}). 
In this case, the flows allocated dedicated per-flow storage within their ingress pipes may still end up using the per-packet model if their storage indices collide when exiting from the same egress pipe. This would lead to the underutilization of storage resources initially reserved for these flows within their ingress pipes.

\parab{Runtime Programmability.} 
The on-switch analysis model of \sys can be programmed in runtime. Specifically, the weights of RNN layers, the escalation thresholds, the number of classification classes, and widths of the inputs and outputs of each layer (\ie the number of binary neurons) are all programmable via the control plane. 
For instance, the weights can be reconfigured by updating the table entries from the control plane.

\parab{On-switch Statistics Collection.} 
To collect the evaluation results from our testbed, we use the second pipe on our switch to implement a result collection module. Specifically, we allocate registers to count the numbers of  escalated packets, packets analyzed by per-packet model, packets analyzed by binary RNN, and pre-analysis packets. Further, we allocate a register array for reporting the on-switch analysis precision and recall for each class, using the combination of ground-truth label and predict label as index. We read these registers from the control plane to obtain the raw data for calculating the macro-F1 scores. 

\parab{Flow Replayer.} 
To generate traffic according to our pcap files, we investigate both \textsf{tcprelay} and DPDK \textsf{pktgen}. 
We choose to use \textsf{pktgen} because it can generate high-throughput traffic that saturates the physical 100 Gbps NIC on our testbed. 
Yet, the key problem of \textsf{pktgen} is that it fails to honor the packet timestamps when sending traffic. However, the on-switch RNN relies on inter-packet delays for inference. 
To work around this issue, we embed the desired timestamp of each packet within the MAC address field of its Ethernet frame. The on-switch analysis pipeline reads this field for flow management and inference. 
We create 32 pcap files throughout the evaluation. When the flow replayer sends an excessively large pcap file that cannot be loaded into the memory at once, it breaks the file into smaller slices and replays these slices sequentially.

\begin{figure*}[t]
    \centering
    \includegraphics[width=\textwidth]{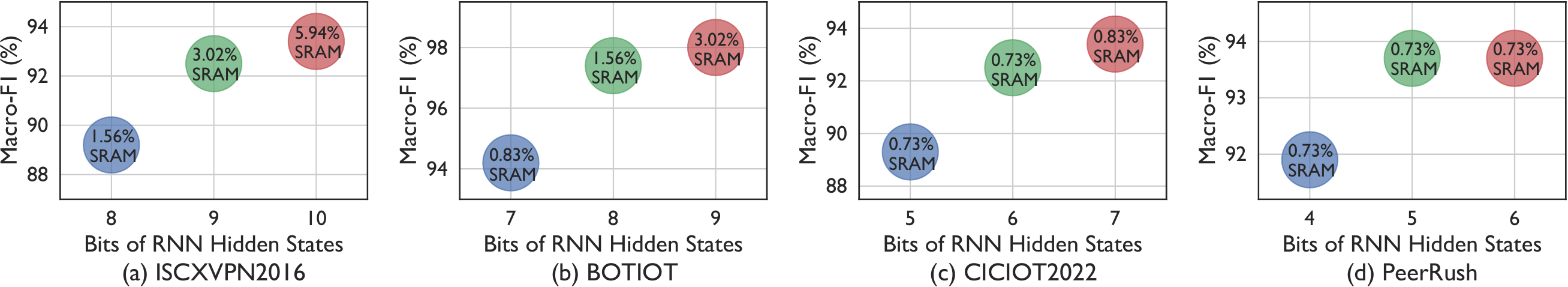}
    \caption{\label{fig:model_size} [Testbed] The traffic analysis accuracy given different binary RNN model sizes.}
\end{figure*}

\parab{Stress Test of Standalone \imis.} 
To stress test the system performance of \imis (\S~\ref{subsec:deepdive}), we generate flows on a server with DPDK packet generator (\textsf{pktgen} version 23.06). These flows are sent directly to the  server where we deploy \imis, bypassing the on-switch analysis. To generate a burst of concurrent flows, the packet generator repeatedly sends packets within a group of selected 5-tuples and the packet size is fixed as 512 bytes.


\subsection{Additional Details about Datasets}
\label{app:datasets}

\paragraph{Data Pre-processing.} For every dataset used in our evaluations, we collect flow records from the original \textsf{pcap} files using the following procedure. \first We collect the original \textsf{pcap} files for each class in the dataset separately, and all the flow records extracted from a \textsf{pcap} file are labelled as the class of this file. \second For each \textsf{pcap} file, we collect the TCP and UDP packets of IPv4, and remove other irrelevant packets, \eg packets of Domain Name System (DNS), Address Resolution Protocol (ARP), Dynamic Host Configuration Protocol (DHCP) and so on. \third We split a clean \textsf{pcap} file by five tuple, and further split packets of the same five tuple into flow records by inter-packet delays. Specifically, if the inter-packet delay between two packets is greater than \mbox{256 ms}, we consider the latter packet as the first packet of a new flow record. This is consistent with our online inference where we consider a flow is completed if we do not receive new packets for the flow for \mbox{256 ms}. \fourth 80\% of flow records in a dataset are used as  the training set and the remaining records are used as testing set.

\paragraph{Traffic Analysis Tasks.} We evaluate \sys using the following tasks.

\begin{itemize}[leftmargin=*]
    \item Encrypted traffic classification on VPN. This task classifies traffic encrypted by Virtual Private Networks (VPNs). We use the ISCXVPN2016~\cite{draper2016characterization} dataset, which contains 7 categories of communication applications captured through the Canadian Institute for Cybersecurity in both VPN and non-VPN. We process the original \textsf{pcap} files for 6 classes of VPN flows, including Email, Chat, Streaming, FTP, VoIP, and P2P. We exclude the Browsing class in our evaluation because some of the applications used for generating Email, Streaming, VoIP packets are web-based, resulting in significant noises, as explained in~\cite{draper2016characterization}. The number of flows in each of the six classes is 613, 2350, 375, 1789, 3495, and 1130, respectively.

    \item Botnet traffic classification on IoT. This task classifies different botnet traffic collected from the Internet of Things (IoT) systems. We process the original \textsf{pcap} files for 4 classes of flows (Data Exfiltration, Key Logging, OS Scan, Service Scan) from the BOTIOT~\cite{koroniotis2019towards} dataset, collected in a realistic network environment deployed in the Cyber Range Lab of UNSW Canberra. The number of flows in each class is 353, 427, 1593, and 7423, respectively.

    \item Behavioral analysis of IoT Devices. This task classifies traffic generated by IoT devices in different working states. We collect the original \textsf{pcap} files for 3 classes (Power, Idle, Interact) from the CICIOT2022~\cite{dadkhah2022towards} dataset, which contains 40 devices of audio, camera, home automation and so on. We process the original \textsf{pcap} files for the Power and Interact classes, and select one day from the 30 days of Idle \textsf{pcap} files. The number of flows in each class is 1131, 4382, and 1154, respectively.

    \item P2P application fingerprinting. This task classifies P2P application traffic. We process the original \textsf{pcap} files for 3 classes (eMule, uTorrent, and Vuze) from the PeerRush dataset~\cite{rahbarinia2013peerrush}. Each class captures one hour of traffic. The number of flows in each class is 20919, 9499, and 7846, respectively.
\end{itemize}

\subsection{Reproducing~\cite{netbeacon} and \cite{N3IC}}
\label{app:reproduce}

We reproduce two recent art NetBeacon~\cite{netbeacon} and N3IC~\cite{N3IC} for evaluation. 

\begin{itemize}[leftmargin=*]
    \item NetBeacon~\cite{netbeacon}: a reproduced version of NetBeacon, which deploys multi-phase tree-based models on switch using both flow-level features and per-packet features. 
    We use the same per-packet features as in~\cite{netbeacon}, and use the max, min, mean, and variance of the packet size and IPD as flow-level features. The inference points are located at the \{$8^\text{th}, 32^\text{nd}, 256^\text{th}, 512^\text{nd}, 2048^\text{th}$\} packet. For each phase we train a 3$\times$7 (3 trees with max depth 7) Random Forest model (their largest model).

    \item N3IC~\cite{N3IC}: a reproduced version of N3IC, which deploys binary MLP on SmartNIC using both statistical flow-level features and per-packet features. 
    We use the same features and phases as NetBeacon for fair comparison, and for each phase the number of neurons in the hidden layers is [128, 64, 10] (their largest model). 
    Note that N3IC deploys binary MLP on SmartNIC but the model cannot be deployed on P4 switches due to hardware resource constraints. 
    Thus, we simulate the switch-side traffic management logic and the binary MLP inference in software to to obtain the traffic analysis results for N3IC. 
\end{itemize}


\subsection{Binary RNN Model Complexity}
\label{app:complexity}

The number of bits allocated to store the RNN hidden states determines both the performance of our RNN model and the size of match-action table for a GRU layer. In Figure~\ref{fig:model_size}, we present the performance of \sys under different bit lengths. The default bit lengths used in four tasks are 9, 8, 6, and 5, respectively. This is because further increasing the bit lengths does not significantly improve F1 scores, while it does increase the SRAM consumption (especially for the first two tasks). When the bit length is smaller than 6, the size of one GRU table is smaller than the minimum allocation unit of SRAM. Thus, further reducing bit lengths will not reduce SRAM consumption.

\end{document}